\documentclass{nature}

\input{epsf}

\usepackage{amssymb}
\usepackage{amsmath}
\usepackage[dvips]{graphicx}
\usepackage{graphics}
\usepackage{txfonts}
\usepackage{xcolor}
\usepackage{times,epsfig}
\usepackage{caption}
\usepackage[utf8]{inputenc}
\usepackage{longtable}

\usepackage{natbib} 
\usepackage{multirow}
\usepackage{multibib}

\newcites{methods}{ }

\bibpunct[, ]{(}{)}{,}{n}{}{,}

\newcommand{\swift}{\textit{Swift~}}
\newcommand{\arcsec}{$''$}

\title{A fast rising tidal disruption event from a candidate intermediate mass black hole}
\author{C.~R.~Angus$^{1}$, V.~F.~Baldassare$^{2}$, B.~Mockler$^{3}$, R.~J.~Foley$^{3}$, E.~Ramirez-Ruiz$^{3,1}$, S.~I.~Raimundo$^{4,1,5}$, K.~D.~French$^{6,7}$, K.~Auchettl$^{3,8,9}$, H.~Pfister$^{1,10}$, C.~Gall$^{1}$, J.~Hjorth$^{1}$, M.~R.~Drout$^{11,12}$, K.~D.~Alexander$^{13}$, G.~Dimitriadis$^{3,14}$, T. ~Hung $^{3}$, D.~O.~Jones$^{3}$, A.~Rest$^{15,16}$, M.~R.~Siebert$^{3}$, K.~Taggart$^{3}$, G. Terreran$^{17,18}$, S.~Tinyanont$^{3}$, C.~M.~Carroll$^{2}$, L.~DeMarchi$^{13}$, N.~Earl$^{6}$, A. Gagliano$^{6,7}$, L.~Izzo$^{1}$, V.~A.~Villar$^{19,20,21}$, Y.~Zenati$^{15}$, N.~Arendse$^{1}$, C.~Cold$^{1}$, T.~J.~L.~de Boer$^{22}$, K.~C.~Chambers$^{22}$, D.~A.~Coulter$^{3}$, N.~Khetan$^{1}$, C.~C.~Lin$^{22}$, E.~A.~Magnier$^{22}$, C.~Rojas-Bravo$^{3}$, R.~J.~Wainscoat$^{22}$ and R.~Wojtak$^{1}$}

\begin{document}

\maketitle

%
\begin{affiliations}
\item DARK, Niels Bohr Institute, University of Copenhagen, Jagtvej 128, DK-2200 Copenhagen \O, Denmark
\item Department of Physics \& Astronomy, Washington State University, Pullman, Washington 99164, USA
\item Department of Astronomy and Astrophysics, University of California, Santa Cruz, CA 95064, USA
\item University of Southampton, Highfield, Southampton SO17 1BJ, UK
\item Department of Physics and Astronomy, University of California, Los Angeles, CA 90095, USA
\item Department of Astronomy, University of Illinois at Urbana-Champaign, 1002 W. Green St., IL 61801, USA
\item Center for Astrophysical Surveys, National Center for Supercomputing Applications, Urbana, IL, 61801, USA
\item OzGrav, School of Physics, The University of Melbourne, Parkville, Victoria 3010, Australia.
\item ARC Centre of Excellence for All Sky Astrophysics in 3 Dimensions (ASTRO 3D)
\item Department of Physics, The University of Hong Kong, Pokfulam Road, Hong Kong, China
\item David A. Dunlap Department of Astronomy and Astrophysics, University of Toronto, 50 St. George Street, Toronto, Ontario, M5S 3H4, Canada
\item The Observatories of the Carnegie Institute for Science, 813 Santa Barbara St., Pasadena, CA 91101, USA
\item Center for Interdisciplinary Exploration and Research in Astrophysics (CIERA) and Department of Physics and Astronomy, Northwestern University, Evanston, IL 60208, USA
\item School of Physics, Trinity College Dublin, The University of Dublin, Dublin 2, Ireland
\item Department of Physics and Astronomy, The Johns Hopkins University, Baltimore, MD 21218, USA
\item Space Telescope Science Institute, Baltimore, MD 21218, USA
\item Las Cumbres Observatory, 6740 Cortona Dr. Suite 102, Goleta, CA, 93117, USA
\item Department of Physics, University of California, Santa Barbara, Santa Barbara, CA, 93106, USA
\item Department of Astronomy \& Astrophysics, The Pennsylvania State University, University Park, PA 16802, USA
\item Institute for Computational \& Data Sciences, The Pennsylvania State University, University Park, PA, USA
\item Institute for Gravitation and the Cosmos, The Pennsylvania State University, University Park, PA 16802, USA
\item Institute for Astronomy, University of Hawaii, 2680 Woodlawn Drive, Honolulu, HI 96822, USA
\end{affiliations}
%
%
\begin{abstract}

Massive black holes (BHs) at the centres of massive galaxies are ubiquitous. The population of BHs within dwarf galaxies, on the other hand, is evasive. Dwarf galaxies are thought to harbour BHs with proportionally small masses, including intermediate mass BHs, with masses $10^{2}<$\,M$_{BH}$\,$<10^{6}$\,M$_{\odot}$. Identification of these systems has historically relied upon the detection of light emitted from accreting gaseous discs close to the BHs. Without this light, they are difficult to detect. Tidal disruption events (TDEs), the luminous flares produced when a star strays close to a BH and is shredded, are a direct way to probe massive BHs. The rise times of these flares theoretically correlate with the BH mass. Here we present AT\,2020neh, a fast rising TDE candidate, hosted by a dwarf galaxy. AT\,2020neh can be described by the tidal disruption of a main sequence star by a $10^{4.7} - 10^{5.9}$M$_{\odot}$ BH. We find the observable rate of fast rising nuclear transients like AT\,2020neh to be rare, at $\lesssim$\,$2 \times10^{-8}$~ events Mpc$^{-3}$~yr$^{-1}$. Finding non-accreting BHs in dwarf galaxies is important to determine how prevalent BHs are within these galaxies, and constrain models of BH formation. AT\,2020neh-like events may provide a galaxy-independent method of measuring IMBH masses.

\end{abstract}
%


In Figure 1 we present the nuclear transient, AT\,2020neh. AT\,2020neh was first reported by the Zwicky Transient Facility \citep[ZTF;][]{Masci2019} on 19th June 2020 at right ascension 15$^{h}$21$^{m}$20.07$^{s}$ and declination +14$^{\circ}$04'10.74'' (J2000), and was confirmed with Young Supernova Experiment data \citep[YSE; ][]{Jones2021}, which showed an initial ealier detection on 17th June 2020. The location of the transient, confirmed in late time imaging from the Hubble Space Telescope (Figure 1), is coincident with the nucleus of the galaxy, lying within 0.1" of the centre. Host-galaxy spectral lines constrain the redshift of the event to $z = 0.062$ ($\sim$280 Mpc). AT\,2020neh reached peak brightness on 1st July 2020, and was monitored with multi-wavelength follow-up observations for over 400 days from peak in the rest frame (for details of the follow-up campaign, see Methods). The full ultraviolet and optical light curves for AT\,2020neh are shown in Extended Data Figure 1.


We present our spectroscopic follow-up observations of AT\,2020neh in Figure 2. The classification spectrum, obtained using the Nordic Optical Telescope on 25th June 2020, 6 days before maximum light, shows a strong blue continuum with a clearly blended helium {\sc{ii}} $\lambda 4685$ and nitrogen {\sc{iii}} $\lambda 4640$ emission feature, and no traces of hydrogen. This blended emission feature has been observed for several optical TDEs \citep{Blagorodnova2019,Leloudas2019,Holoien2020,vanVelzen2021,Charalampopoulos2022}, and is attributed to a fluorescence mechanism requiring both a high-energy radiation source and a high gas density \citep{Bowen1934}. Given the nuclear location of the transient (Figure 1), we interpret these features under a TDE classification for the transient AT\,2020neh. The spectra become featureless after maximum light, evolving to gradually reveal a broad hydrogen, H$\alpha$, emission line at +36~d (and later in H$\beta$ too). This H$\alpha$ emission dominates the late-time spectra of AT\,2020neh, exhibiting an asymmetric profile which is blueshifted with respect to the rest frame by $\sim$4000~km~s$^{-1}$ (see Extended Data Table 1). The profile of this emission is consistent with emission lines arising from optically thick outflowing material \citep{RothKasen2018}, which has been seen in several other TDEs \citep[e.g.][]{Blanchard2017,Hung2019,Nicholl2020,Charalampopoulos2022}. The lack of elements heavier than hydrogen within the late time ($>$200~d) spectra is consistent with a TDE classification, as these elements would only be expected if the event arose from a stellar explosion \citep{Jerkstrand2017J}.


Only a dozen TDEs have optical light curves with sufficient data coverage at early times to confidently determine the time taken for the transient to reach maximum light \citep[see][]{Gezari2021}. In Figure 3 we show the distribution of their rise times and luminosities, alongside those of AT\,2020neh. The constraining non-detections from frequent YSE monitoring prior to first detection of the transient, and the depth of the YSE imaging, result in an exceptionally well constrained rise time for this event. With information on the evolution of AT\,2020neh down to 3\% of its maximum light, we measure a rise time from initial detection to $g$-band peak of only $13.2\pm1.0$~days in the rest frame. To date, this is the fastest measured rise time for a TDE, on average a factor of $2.4 ^{+3.0}_{-1.5}$ times faster than others with pre-max light curve coverage. AT\,2020neh attained a peak bolometric luminosity of $ (4.2\pm0.1)\times10^{43}$ erg s$^{-1}$. Whilst this luminosity alone is not unprecedented amongst TDEs, when taken into consideration with its short rise time, it places AT\,2020neh on the periphery of the known nuclear transient luminosity-rise parameter space, being brighter than other confirmed TDEs with short rise times. This unique location places AT\,2020neh between TDEs and another class of transients - namely \lq Fast and Blue Optical Transients\rq~(FBOTs) \citep[e.g.][]{Vinko2015, Prentice2018,Pursiainen2018}. Despite this similarity in rise time, AT2020neh shows significant differences in both its spectroscopic properties and its photometric evolution from those of FBOTs (see Methods), which set it apart from this class. We find that the post-maximum evolution of AT\,2020neh is typical of other TDEs, being well described by a power-law decline slope of $\alpha=1.44 \pm 0.19$. This is consistent with the theoretical $-5/3$ slope expected from transients powered by fallback accretion \citep{Rees1988}.


The host galaxy of AT\,2020neh, SDSSJ152120.07+140410.5, is unusual amongst the population of TDE host galaxies (see Methods). The galaxy is blue in color and a pre-explosion SDSS spectrum of the host galaxy shows that it is star forming (see Extended Data Figure 4). However, the galaxy also has a high S\'{e}rsic index and a high surface density concentration of stellar mass relative to field galaxies. This increases the number of stars capable of being disrupted, which has been seen within other TDE host galaxies \citep{LawSmith2017,Graur2018,French2020}. When we measure the stellar mass of SDSSJ152120.07+140410.5, we find it to be $\log(M_{\star}/M_{\odot}) = 9.57 \pm 0.20$. This mass is consistent with the masses of dwarf galaxies, and is about the same as the mass of the Large Magellanic Cloud. Scaling relations between galaxy stellar mass and BH mass (derived from more massive galaxy populations, \citep{ReinesVolonteri2015}) predict that for such small galaxies, we expect correspondingly low-mass BHs. Direct evidence for the presence of BHs in low-mass galaxies however, is scarce \citep[see][]{Reines2016}. These systems are typically faint, and the gravitational sphere of influence for low-mass BHs is predicted to be small, making them difficult to resolve in observations of galaxies outside of the Local Group. Thus the occupation fraction of BHs at the low mass end of the galaxy stellar mass regime remains observationally unconstrained \citep{Miller2015}. The detection of AT\,2020neh thus indicates the presence of a BH in a dwarf galaxy, from which we can begin to explore the properties of the BH.


The unique location of AT\,2020neh in Figure 3 implies a relative scarcity of similar events like this amongst the TDE population. Indeed, we estimate the observable rate of fast rising TDEs, like AT\,2020neh to be $\lesssim${}$2\times10^{-8}$~Mpc$^{-3}$~yr$^{-1}$ at $z=0.27$, the detection limit of the YSE survey (see Methods). This is approximately one twenty-fifth of the `normal' TDE rate \citep{vanVelzen2020}. The relatively low rate of fast TDEs may be explained through the mechanics of the BH-stellar encounter. For low ($<10^{6}$M$_{\odot}$) mass BHs, the event horizon of the BH is much smaller than the tidal radius required to disrupt a star of average mass. Therefore in order for the event to be observed as a `classical TDE', one whose luminosity closely follows the rate at which material falls onto the BH (the `fallback rate'), a plunging encounter is required. Such an encounter allows the material to circularise quickly enough for the event to be observable \citep{Guillochon2014}, as found when modelling AT\,2020neh (see Methods). Events with more shallow, grazing encounters are unlikely to circularise their debris into an accretion disc on timescales shorter than the timescale of peak fallback accretion. This is a consequence of weaker general relativistic effects around smaller BHs. These encounters are thought to produce less luminous transients with irregular light curves which are unlikely to be observable \citep{Guillochon2015}. We estimate the peak luminosity of AT\,2020neh to exceed the maximum (Eddington) luminosity at which the BH can radiate by a factor of $\sim${}$(2-3) \times L_{\rm Edd}$ for its measured BH mass range \citep{Dai2018}. This is in agreement with theoretical predictions, as observable TDEs around lower mass BHs are also likely to be `super-Eddington' flares, as a result of their predicted high fallback rates \citep{Guillochon2015}.


The properties of the TDE light curve can potentially be used to probe the BH mass \citep[][]{Mockler2019}, which could provide a measurement independent of assumptions about the host galaxy. The rise time of a TDE, $\Delta t$, which is the time taken between disruption of the star to peak luminosity of the event, theoretically correlates with the BH mass. This is because the luminosity of the transient is expected to follow the fallback rate of the stellar debris, with a relationship $\Delta t \propto {M_{\rm BH}}^{0.5}$ \citep[][see Methods for modelling details]{Mockler2019}. Fitting the multiband UV-optical light curve of AT\,2020neh (Extended Data Figure 6) we find $\log{M_{\rm BH}/M_{\odot}}=5.5^{+0.4}_{-0.3}$. We note that the physical origins of the early optical emission in TDEs is unknown, so this approach to BH mass estimation may not be applicable to all events. We assess the reliability of this measurement with two galaxy-BH scaling relationships. The $M_{\star} - M_{\mathrm{\rm BH}}$ scaling relation \citep{ReinesVolonteri2015} predicts a BH mass of $\log{M_{\rm BH}/M_{\odot}} = 5.9 \pm 0.2$. We use a late-time high-resolution spectrum of the TDE to measure the velocity dispersion ($\sigma$) of the host galaxy. We measure $\sigma=39\pm$13 km s$^{-1}$, which using an $M_{\mathrm{BH}} - \sigma$ relation \citep{MerrittFerrarese2001} provides a BH mass of $\log{M_{\rm BH}/M_{\odot}} = 4.8 ^{+0.5}_{-0.9}$. These are both consistent with the mass estimated from the rise of the light curve. Our constraints on the mass range of the BH, $ 4.7 < \log{M_{\rm BH}/M_{\odot}} < 5.9$, place it within the domain of `intermediate mass BHs' (IMBHs), whose masses span the gap between stellar-mass and supermassive BH populations ($2 \lesssim \log{M_{\mathrm{BH}}/M_{\odot}} \lesssim 6$), but for which evidence of a firm population has been difficult to obtain \citep{Greene2020}.


The massive BHs ($M_{BH}\gtrsim 10^{9}M_{\odot}$) we observe in the hearts of galaxies in the local Universe \citep[][]{Magorrian1998} must originate from less massive BHs which have grown through mergers or accretion of surrounding material. How this putative \lq seed\rq\; population of less massive BHs forms is unclear. Possible formation mechanisms include hierarchical mergers of stellar-mass BHs \citep{Giersz2015}, core-collapse of exceptionally massive stars in the early Universe \citep{Schneider2002}, and the direct collapse of massive gas clouds \citep{Loeb1994}. Measuring the low-mass ends of the BH mass function and BH-galaxy scaling relationships is important for distinguishing between these scenarios, as each model is predicted to influence the slope of these relationships differently \citep{Greene2020}. Identifying BHs in dwarf galaxies and measuring their masses is generally difficult \citep{Reines2016,Baldassare2017,Baldassare2018,Reines2020} due to their intrinsically low luminosity and smaller gravitational influence. Typically, signatures of BH accretion (an Active Galactic Nucleus, `AGN') are required to reliably confirm the presence of a BH in a dwarf galaxy \citep[e.g.][]{Kunth1987,Filippenko1989,Baldassare2015,Baldassare2020}. However, the majority of BHs are not active. Fast-rising optical TDE candidates such as AT\,2020neh therefore may offer an opportunity to find and study non-active BHs in dwarf galaxies.


In Figure 4 we display the location of AT\,2020neh on two commonly used BH-galaxy scaling relations. AT\,2020neh is the first candidate optical TDE from an IMBH in a dwarf galaxy. Moreover, its host galaxy is one of only a small number of dwarf galaxies which has a BH mass estimate independent of galaxy scaling relationships. At present, most galaxies in the low-stellar mass/low-velocity dispersion regions of BH scaling relations are low-luminosity AGN, and therefore not representative of the quiescent dwarf galaxy population. A handful of X-ray detected transients in dwarf galaxies have also had a IMBH-TDE origin proposed \citep{Donato2014,Lin2018,He2021}, for which the BH mass estimates have been derived from galaxy-scaling relationships from fitting of the X-ray spectrum with disc models. AT\,2020neh represents the first of a population of optically detected transients whose properties  may allow us to explore the low-mass end of BH-scaling relations in a complementary way. The relative rate of fast-TDEs to normal-TDEs implies that other events like AT\,2020neh should be more common amongst the TDE population. Early detection of future fast TDEs within deep, high-cadenced data sets, like the early YSE detection of AT\,2020neh, will generate the samples of events with constraining data sets needed to test the reliability of estimating BH masses from rise-time measurements. The properties of AT\,2020neh will form the baseline from which targeted observing strategies can be designed to ensure maximum scientific gain from the discovery of fast TDEs in current and future surveys. We estimate that YSE will observe another 5-6 similar events over its lifetime, each with the necessary light curve data to constrain the rise time. 


%
\newpage
%
%
\section*{Methods Summary}

We describe the data, methods and theoretical calculations used within this work. We outline the different photometric and spectroscopic data sets used and describe the reduction and analysis techniques adopted. We discuss the interpretation of the transient as a TDE against other spectroscopic classes, and provide details of the localization of AT\,2020neh within its host galaxy. We present the analysis used to determine the host-galaxy properties and the different methods used to infer the BH mass. Finally, we present our methodology for the determination of the rate of AT\,2020neh-like events, and estimate the number of similar events which could be identified in current and next-generation surveys. Throughout this work we assume a standard $\Lambda$~CDM cosmology with H$_{0}$=71 km s$^{-1}$ Mpc$^{-1}$, $\Omega_{M}$=0.27 and $\Omega_{vac}$=0.73.

%
\newpage

\bibliographystyle{naturemag}
\bibliography{main}

\section*{Author information}
Reprints and permissions information is available at
www.nature.com/reprints. The authors declare no 
competing financial interests. Readers are welcome to 
comment on the online version of the paper. Correspondence
and requests for materials should be addressed to 
C.R.A. (angus@nbi.ku.dk).

%
\section*{Acknowledgments}
We thank M. Pursiainen for insightful discussions regarding the spectroscopic evolution of this event, and M. Briday for useful instructions regarding the use of {\sc{Prospector}}.

C.R.A., J.H., K.A., L.I., C.C., N.K. and R.W. were supported by VILLUM FONDEN Investigator grant (project \# 16599). C.R.A., C.G. and L.I. were supported by a VILLUM FONDEN Young Investigator Grant (project \#25501). B.M. acknowledges support from the AAUW Dissertation Fellowship, Swift grant 80NSSC21K1409. The UCSC team is supported in part by NASA grant 80NSSC20K0953, NSF grant AST--1815935, the Gordon \& Betty Moore Foundation, the Heising-Simons Foundation, and by a fellowship from the David and Lucile Packard Foundation to R.J.F.~ E.R.-R. acknowledges support by Heising-Simons Foundation, Nasa Swift and NICER and NSF (AST-1911206 and AST-1852393). S.I.R. has received funding from the European Union’s Horizon 2020 research and innovation program under the Marie Sklodowska-Curie grant agreement No. 891744. Parts of this research were supported by the Australian Research Council Centre of Excellence for All Sky Astrophysics in 3 Dimensions (ASTRO 3D), through project number CE170100013 and the Australian Research Council Centre of Excellence for Gravitational Wave Discovery (OzGrav), through project number CE170100004. H.P. acknowledges support from the Danish National Research Foundation (DNRF132) and the Hong Kong government (GRF grant HKU27305119, HKU17304821). M.R.D acknowledges support from the NSERC through grant RGPIN-2019-06186, the Canada Research Chairs Program, the Canadian Institute for Advanced Research (CIFAR), and the Dunlap Institute at the University of Toronto. D.O.J. is supported by NASA through the NASA Hubble Fellowship grant HF2-51462.001 awarded by the Space Telescope Science Institute, which is operated by the Association of Universities for Research in Astronomy, Inc., for NASA, under contract NAS5-26555. M.R.S. is supported by the National Science Foundation Graduate Research Fellowship Program under grant No. 1842400. D.A.C. acknowledges support from the National Science Foundation Graduate Research Fellowship under Grant DGE1339067. A.G. is supported by the National Science Foundation Graduate Research Fellowship Program under Grant No.~DGE–1746047. A.G. also acknowledges funding from the Center for Astrophysical Surveys Fellowship at UIUC/NCSA and the Illinois Distinguished Fellowship.

Some of the data presented herein were obtained at the W. M. Keck Observatory, which is operated as a scientific partnership among the California Institute of Technology, the University of California, and NASA. The Observatory was made possible by the generous financial support of the W. M. Keck Foundation. The authors wish to recognize and acknowledge the very significant cultural role and reverence that the summit of Maunakea has always had within the indigenous Hawaiian community. We are most fortunate to have the opportunity to conduct observations from this mountain.

Based on observations made with the Nordic Optical Telescope, owned in collaboration by the University of Turku and Aarhus University, and operated jointly by Aarhus University, the University of Turku and the University of Oslo, representing Denmark, Finland and Norway, the University of Iceland and Stockholm University at the Observatorio del Roque de los Muchachos, La Palma, Spain, of the Instituto de Astrofisica de Canarias. Observations were carried out under program P61-022.

A major upgrade of the Kast spectrograph on the Shane 3~m telescope at Lick Observatory was made possible through generous gifts from the Heising-Simons Foundation as well as William and Marina Kast. Research at Lick Observatory is partially supported by a generous gift from Google.

This research is based on observations made with the NASA/ESA Hubble Space Telescope obtained from the Space Telescope Science Institute, which is operated by the Association of Universities for Research in Astronomy, Inc., under NASA contract NAS 5--26555. These observations are associated with program SNAP-16239.

We acknowledge the use of public data from the Swift data archive.

Based on observations obtained at the international Gemini Observatory (program GN-2020A-DD-111), a program of NSF’s NOIRLab, which is managed by the Association of Universities for Research in Astronomy (AURA) under a cooperative agreement with the National Science Foundation. on behalf of the Gemini Observatory partnership: the National Science Foundation (United States), National Research Council (Canada), Agencia Nacional de Investigaci\'{o}n y Desarrollo (Chile), Ministerio de Ciencia, Tecnolog\'{i}a e Innovaci\'{o}n (Argentina), Minist\'{e}rio da Ci\^{e}ncia, Tecnologia, Inova\c{c}\~{o}es e Comunica\c{c}\~{o}es (Brazil), and Korea Astronomy and Space Science Institute (Republic of Korea). We thank the Director for supporting this program.

Based in part on observations obtained with the Samuel Oschin 48-inch Telescope at the Palomar Observatory as part of the Zwicky Transient Facility project. ZTF is supported by the NSF under grant AST-1440341 and a collaboration including Caltech, IPAC, the Weizmann Institute for Science, the Oskar Klein Center at Stockholm University, the University of Maryland, the University of Washington, Deutsches Elektronen-Synchrotron and Humboldt University, Los Alamos National Laboratories, the TANGO Consortium of Taiwan, the University of Wisconsin at Milwaukee, and the Lawrence Berkeley National Laboratory. Operations are conducted by the Caltech Optical Observatories (COO), the Infrared Processing and Analysis Center (IPAC), and the University of Washington (UW).

The Pan-STARRS1 Surveys (PS1) and the PS1 public science archive have been made possible through contributions by the Institute for Astronomy, the University of Hawaii, the Pan-STARRS Project Office, the Max-Planck Society and its participating institutes, the Max Planck Institute for Astronomy, Heidelberg and the Max Planck Institute for Extraterrestrial Physics, Garching, The Johns Hopkins University, Durham University, the University of Edinburgh, the Queen's University Belfast, the Harvard-Smithsonian Center for Astrophysics, the Las Cumbres Observatory Global Telescope Network Incorporated, the National Central University of Taiwan, STScI, NASA under grant NNX08AR22G issued through the Planetary Science Division of the NASA Science Mission Directorate, NSF grant AST-1238877, the University of Maryland, Eotvos Lorand University (ELTE), the Los Alamos National Laboratory, and the Gordon and Betty Moore Foundation.

%
%
\section*{Author contributions}

C.R.A. led the overall project and analysis, and wrote the majority of the paper. They are also PI of the NOT and Gemini programs used to collect data. V.F.B. performed the analysis to searched for previous variability of host galaxy, wrote text, made the M-$\sigma$ figure, and contributed to discussions regarding the link between the BH and host galaxy. B.M. performed the MOSFiT light curve analysis, made the MOSFiT figure and wrote text. R.J.F. reduced the HST data and observed for the Keck nights. They are PI of Keck, Lick, and Hubble programs used to collect data. They also wrote text, contributed to contributed to discussions regarding interpretation of the object and provided feedback on the manuscript. E.R.-R. contributed to discussions regarding the theoretical interpretation of the event.  S.I.R. helped to perform the pPXF fitting of the host galaxy features and contributed to discussions. K.D.F. contributed to discussions regarding the nature of host galaxy and provided extensive feedback on manuscript drafts. K.A. triggered, reduced and analysed the {\textit{Swift}} data used, contributed to discussions of results and observations. They also wrote text for the paper and provided comments on the manuscript. H.P. performed part the rate analysis of fast TDEs, contributed to discussions regarding the interpretation of the event and provided feedback on manuscript drafts. They also performed vetting of candidates within the YSE survey. C.G. performed spectral line profile analysis of the H$\alpha$ emission, contributed to discussions and provided feedback on manuscript drafts. J.H. contributed to discussions and provided extensive feedback on manuscript drafts.  M.R.D. contributed to discussions and performed initial analysis regarding the rate of fast tidal disruption events within YSE. K.D.A. triggered and reduced radio observations with the VLA, and contributed to the interpretation of the resulting upper limits. G.D. reduced the GMOS spectra. T.H. reduced data, observed and submitted observing proposals for the event. D.O.J. reduced the YSE data for the event and performed PS1 operations for the survey. A.R. analysed and helped to interpret the forced YSE photometry of the event. M.R.S. reduced the LRIS and KAST spectra. K.T. provided manuscript comments and observed with Keck. G.T. observed and reduced the DEIMOS spectrum from Keck. C.~M.C., L.~DeM. N.E. and A.G. provided extensive feedback for the manuscript as a junior review panal. L.I., S.T., V.A.V., Y.Z. also provided comments on the paper. N.A., C.C., N.K. and R.J.W. performed vetting of candidates within the YSE survey. D.A.C. wrote the collaboration software platform YSE-PZ and support it and the server it runs on. T.J.L.de B., K.C.C., C.C.L., E.A.M. and C.R-B. perform PS1 operations, including supporting the Pan-STARRS IPP which provides the initial astrometry and photometry. C.R.A., R.J.F., S.I.R., K.A., H.P., J.H., C.G., L.I., N.A., C.C. and R.W. all aperformed vetting of candidates within the YSE survey.

%
\newpage
\begin{methods}

\section*{Data}

\subsection{{\textit{Optical Photometry}}}

The optical photometry of AT\,2020neh is obtained by the Young Supernova Experiment (YSE) sky survey \citep{Jones2021} with the 1.8m Pan-STARRS telescope (PS1). PS1 is mounted with a 1.4 Gigapixel camera \citemethods{Kaiser2002} to image the sky in the $g$, $r$, $i$, and $z$ filters. The data are downloaded, processed and archived using the Image Processing Pipeline (IPP) at the University of Hawaii’s Institute for Astronomy \citemethods[][]{Magnier2020}. Images from the PS1 3$\pi$ survey are used as references for template subtraction, with each reference image convolved to match the point spread function (PSF) of the nightly observations, before passing through the Transient Science Server at Queens University Belfast \citemethods[][]{Smith2020} to identify new events. The YSE photometric pipeline is based on {\sc{photpipe}} \citepmethods{Rest2005}. Forced PSF photometry is performed for each transient, using a flux weighted centroid matching the PSF at the transient location.

The field in which AT\,2020neh is located was monitored by YSE for 3 months prior to the first detection of the transient, with non-detections at 18 epochs over this period, in which no light from the transient is observed. These data provide strong constraints upon the explosion epoch down to limiting magnitudes of typically $m\approx21.5$~mag in the $g$, $r$ and $i$ bands and $m\approx20.5$~mag in $z$.  Additional $g$- and $r$-band photometry of AT\,2020neh is taken from the public Zwicky Transient Facility \citepmethods[ZTF, ][]{Bellm2019,Graham2019} data stream, with observational coverage from 19 June 2020 to 04 August 2020 ($-12$ to +35 days in the rest frame phase).

\subsection{{\textit{Optical Spectroscopy}}}

Spectra were obtained using the instruments and observational set-ups listed in Extended Data Table 2. The ALFOSC, Kast and LRIS spectra were reduced using custom made pipelines and standard routines within {\sc{iraf}}. The GMOS spectra were  processed using the Gemini {\sc{iraf}} package. The DEIMOS spectrum was processed with the {\ttfamily{PypeIt}} software package \citepmethods{Prochaska2020}. All calibrations and correction procedures are performed after the basic pipeline reduction using custom {{\tt python}} programs. Spectra are corrected for an average Galactic extinction along the line of sight to the transient of $E(B-V) = 0.038$~mag and $R_{V} = 3.1$ based on the dust maps of \citemethods[][]{Schlafly2011}. We use a pre-transient Sloan Digital Sky Survey \citepmethods[SDSS; ][]{Aihara2011} spectrum of the host galaxy to correct for the host-galaxy contribution to the spectra. We color correct both transient and host-galaxy spectra using the technique of \citemethods{Hsiao2007} to the available $g$,$r$,$i$,$z$ photometry from PS1 to correct for slit/fibre losses. For the low-resolution ALFOSC and Kast spectra, we convolve this host-galaxy spectrum to the resolution of these instruments before subtraction. For higher-resolution spectra, we create models of the host-galaxy emission lines based on measurements from the SDSS spectrum and subtract these models from the science spectra. All line measurements are performed using the {\tt specutils} package in {\tt python}\footnote{https://specutils.readthedocs.io/en/stable/index.html}. We note that several narrow stellar absorption features with widths smaller than the instrumental resolution of the SDSS spectrograph still persist within our host-galaxy subtracted spectra (a consequence of the low velocity dispersion of the host galaxy - see below).

\subsection{{\textit{Swift Observations}}}

We requested, and were awarded, target-of-opportunity (ToO) observations from the Neil Gehrels Swift Gamma-ray Burst Mission (\swift) UltraViolet and Optical Telescope (UVOT) and X-ray Telescope (XRT). AT\,2020neh was monitored with \swift over two epochs whilst the transient is active; from 01 Jul 2020 to 17 Jul 2020, and from 05 Aug 2020 to 06 Sept 2020 with an approximate 2 day cadence. Final \swift observations were also obtained between 13 January 2021 to 27 January 2021 and averaged to estimate the host-galaxy level for subtraction. To obtain the UVOT photometry, we used the task {\sc{uvotsource}} with source radii of 5.0 \arcsec and background radii of 40.0 \arcsec. The transient is clearly detected with UVOT during the first two epochs. The host-galaxy subtracted UVOT light curve is presented in Extended Data Figure 1. 

XRT was operated in the Photon Counting mode for all observations. We reprocessed the data with the task {\sc{xrtpipeline}} version 0.13.2, using standard filters, screen and recent calibration files. Using a source region centered on the location of AT\,2020neh with a radius of 49 \arcsec and a 150 \arcsec radius source free background region centered at 15$^{h}$21$^{m}$04.77$^{s}$, +13$^{\circ}$59'37.78'', we detect no significant X-ray emission arising from the source. To constrain any X-ray emission, we merge all 37 Swift XRT observations of AT\,2020neh using the task {\sc{xselect}}. We derive a $3\sigma$ upper limit to the 0.3-10.0 keV count rate of 0.002 counts s$^{-1}$. Assuming an absorbed powerlaw model with a column density of $2.7\times10^{20}$ cm$^{-2}$ \citepmethods{HI4PICollab2016} and $\Gamma=2.7$ (similar to that of other X-ray bright TDEs \citepmethods[e.g.,][]{Auchettl2017}), redshifted to the location of the host galaxy, we find an $3\sigma$ upper limit to the unabsorbed X-ray luminosity of $4.5 \times 10^{41}$ erg s$^{-1}$. This faint X-ray luminosity is consistent with other previously observed TDEs \citepmethods{Auchettl2017} and suggests the source does not harbour a strong AGN component, congruous with the composite location of the host galaxy in the BPT diagram (Extended Data Figure 4).

\subsection{{\textit{Radio Observations}}} AT\,2020neh was observed twice during its evolution with the Very Large Array (VLA) under program VLA/20B-377 (PI: Alexander). The  first observation was conducted on 30 Jun 2020 (-1 days from peak in the TDE restframe) at 15 GHz. The source was not detected with a 3$\sigma$ upper limit of 18 $\mu$Jy. The second observation on 31 Dec 2020 (+172 days restframe) at 6 GHz also resulted in a non-detection, with a 3$\sigma$ upper limit of 16 $\mu$Jy. These limits are consistent with a lack of a relativistic jet \citepmethods{Alexander2020}, and further support the conclusion that the host galaxy does not harbour an obvious AGN (see Host Galaxy Properties).

\subsection{{\textit{HST Observations}}} We observed AT\,2020neh on 16 Sept 2021 with the Wide Field Camera 3 on the {\it Hubble Space Telescope} (HST) under program SNAP-16239 (PI:Foley). AT\,2020neh was imaged in F225W and F275W for a total exposure time of 780 and 710~s, respectively. Two images were taken in each band to reject cosmic rays. The data were processed through the STScI data reduction pipeline, including basic processing, calibration, and drizzling.

AT\,2020neh is clearly detected in the {\it HST} images at the centre of its host galaxy surrounded by a circumnuclear ring with a projected radius of 0.35-0.7'', corresponding to physical distances of 0.45-0.9 kpc at $z = 0.062$. We perform photometry using the automatic detection and extraction package, Source Extractor \citepmethods[SExtractor,][]{Bertin1996}, applying a surface brightness signal-to-noise cut of two per pixel to include faint surface brightness features and adjusting the extraction parameters to separate the TDE from the star forming ring. Zeropoints for each filter were taken from the STScI WFC3 handbook \citepmethods[][]{Dressel2019}. We measure $m_{\mathrm{F225W}}= 22.059 \pm 0.022$ and $m_{\mathrm{F275W}}= 21.790 \pm 0.019$.

\section*{Transient Classification}

The photometric properties of AT\,2020neh make it exceptional amongst the nuclear transient population, with a fast rise and luminous peak, which require careful classification. In particular it is important to distinguish this event from hydrogen rich core collapse SNe (SNe II), as the evolutionary timescales and hydrogen spectral features are similar to this class. It is also important to establish whether AT\,2020neh could be classified as an FBOT event. Below we present the evidence for a TDE origin for AT\,2020neh.

\subsection{\textit{Spectroscopic Behaviour}}

The dominant feature present within the NOT spectrum taken pre-maximum light, is a strong blended He II and N III emission feature around 4660~\AA.  This feature is no longer detectable within the first post-max spectrum obtained at +9d with GMOS. Such short-lived features have been observed within samples of young SNe II, where the high-energy ultraviolet emission which occurs during shockbreak-out briefly ionises any surrounding circumstellar material, recombining to produce strong emission lines \citepmethods[e.g.,][also see Extended Data Figure 2]{Khazov2016}. SNe which exhibit these ``flash ionisation'' features are typically accompanied by Balmer emission lines, which we do not observe with AT\,2020neh at this epoch (all Balmer features observed in the NOT spectrum are consistent with the SDSS spectrum of the host galaxy). Based on the velocities observed for the He II/N III emission ($950 \pm 160$~km~s$^{-1}$), we should have observed any flash ionised Balmer emission within this early spectrum.

The later spectroscopic evolution of AT\,2020neh also does not follow the canonical behaviour of a SN II. We only begin to observe H$\alpha$ emission at later epochs ($>$ 37 days from peak), when it presents a broad, boxy blueshifted emission profile with no accompanying absorption. Whilst there is diversity in the H$\alpha$ features present within SN II spectra \citepmethods[][]{Gutierrez2014,Gutierrez2017}, H$\alpha$ nominally takes on a P-Cygni profile for these events, due to the expanding, optically thick material ejected from the progenitor. For AT\,2020neh to be a SN II, the lack of P-Cygni signature in its H$\alpha$ emission would require either the photosphere to be very small compared to the expanding material, or there to be optically thin material along the line of sight masking the photosphere. The first scenario is rare, seen only in a few SNe-II at very early times \citemethods[]{Gutierrez2014}. The second scenario requires the observed H$\alpha$ emission to be produced via interation (as for SNe IIn), for which narrow we would expect to see very narrow features ($\sim$200 km s$^{-1}$). We do not observe this narrow emission in AT\,2020neh. Though blueshifted Balmer features have been observed within some SNe II, they typically occur at much smaller velocity shifts (population mean $2000$~km~s$^{-1}$, \citemethods[][]{Anderson2014}) compared to the $\sim$4000~km~s$^{-1}$ shift observed within AT\,2020neh. SNe II blueshifts also evolve with time, disappearing by the nebular phase. The H$\alpha$ line remains blueshifted until late times, and does not evolve significantly during the later states of its evolution. Finally, the late time ($>$200 day) spectra of AT\,2020neh do not show the forbidden lines [Ca II], [O I] and [Fe II] traditionally observed during the nebular phase in SNe II \citepmethods[][]{Gutierrez2017,Jerkstrand2017J}. 

There are few FBOTs with spectroscopic data sets sufficient to determine what the typical spectroscopic features of this class are. Those which do generally exhibit hot, featureless spectra \citepmethods{Drout2014,Vinko2015,Pursiainen2018}. The luminous and local AT\,2018cow \citepmethods{Prentice2018,Perley2019} is the only FBOT with sufficient spectroscopic data to make any meaningful comparisons. Post-peak H$\alpha$ emission has also been observed within this event, however it is significantly narrower than the H$\alpha$ seen in AT\,2020neh at similar phase \citepmethods{Perley2019}, even when accounting for the narrower component we see observe at much later epochs (which exhibits velocities of order 2000 km s$^{-1}$).

Though not typical of optically selected TDEs \citepmethods{vanVelzen2020,Hammerstein2022}, these spectroscopic properties can all be interpreted within a TDE framework. We attribute the blended He II and N III early emission feature to Bowen fluorescence (Extended Data Figure 2), which has been observed within several TDEs \citep{Leloudas2019,Charalampopoulos2022}. Blue-shifted, boxy Balmer emission profiles have also been observed within some TDEs \citep[e.g.][]{Hung2019}, and have been shown to result from strongly irradiated, outflowing, optically thick material from the system \citep{RothKasen2018}. Though the widths of the emission lines observed in AT2020neh are narrower than those seen in other TDEs \citep{vanVelzen2020}, these features are viewing angle dependent, and therefore could be a result of a system with low inclination \citepmethods{Parkinson2022}. The lack of nebular phase features is also consistent with a non-supernova origin.

\subsection{\textit{Photometric Behaviour}}

Though the fast rise of AT\,2020neh does coincide with the distribution of rise times observed for SNe II \citepmethods[][]{GonzalezGaitan2015}, it is significantly more luminous, with a peak luminosity four times brighter than the upper end of the SN II luminosity function \citepmethods[][]{Patat1994}. To construct the bolometric light curve of AT\,2020neh, we interpolate the available photometry (corrected for foreground extinction) in each band using gaussian processes, a non-parametric interpolation method which treats all data as though they were drawn from an Gaussian distribution over the possible functions the light curve could take. We implement this using a python package \citemethods[][]{George2014}, interpolating at a daily cadence using a Matern-3/2 kernel. We then fit the resulting spectral energy distributions at each epoch of the interpolated light curves with a simple black body function. From this constructed bolometric curve, we estimate the peak luminosity to be $ 4.2\pm0.1\times10^{43}$ erg s$^{-1}$.

In the luminosity-rise time parameter space shown in Figure 3, AT\,2020neh occupies a unique position, being fast and also brighter than other shorter-lived TDEs. These two properties are comparable to those observed within FBOT events, known for their short-lived optical lifetimes and often bright peak luminosities. FBOTs are also frequently characterised by their rapidly cooling photospheres \citepmethods{Prentice2018,Pursiainen2018,Ho2020}, with some events cooling by $>10,000$K in the early stages of their declines \citepmethods[][]{Prentice2018}. The UV-optical light curve of AT\,2020neh exhibits significant cooling during the first 30 days from peak, with a black body temperature decrease of approximately 9000~K (Extended Data Figure 3). Although a cooling photosphere is not associated with many optically discovered TDEs, the majority of which exhibit a constant or even increasing temperature after peak \citepmethods[][]{Hammerstein2022}, apparent cooling of the photosphere has been observed a few events, e.g. iPTF16fnl, AT\,2019qiz and several TDEs discovered by ZTF \citepmethods{Blagorodnova2019,Nicholl2020,Hinkle2021,Hammerstein2022}. The rate of cooling of AT\,2020neh is comparable to these events, which typically display a $\Delta$T/$\Delta$t $\gtrsim$ 5000 K / 20 days \citepmethods[][]{Nicholl2020,Hinkle2021,Hammerstein2022}, but much slower than the cooling observed in FBOTs like AT\,2018cow (Extended Data Figure 3). As the exact source of the optical emission in TDEs is unknown, alongside the physical mechanisms which produce it, it is not clear why some TDEs exhibit this cooling behaviour, although it has been suggested that it may be linked to optically thick outflows, where cooling of the photosphere occurs when the outflow reaches a characteristic photon trapping radius from which photons are able to escape \citep{Nicholl2020}. However it is currently not clear whether this scenario is applicable to the majority of cooling cases in TDEs.

From only photometric information, it is difficult to discern FBOTs from TDEs from lower mass BHs. The physical origins of FBOTs are disputed, but some have had a TDE origin suggested for their production \citepmethods[e.g.][]{Vinko2015,Perley2019}. We use the transient rise times to estimate approximate black hole masses for TDEs (shown along the upper axis of Figure 3), assuming a standard encounter with a solar mass star. Using this metric, we can see that \lq faster rising\rq~ AT\,2020neh-like events probe the regions of super-Eddington accretion for lower (M$_{BH}<10^{5.5}$M$_{\odot}$) mass BHs.

\subsection{\textit{Astrometric Location}}

The location of AT\,2020neh within its host galaxy can also be used to help discriminate between theories for its origin. AT\,2020neh is located in SDSSJ152120.07+140410.5, a galaxy at a spectroscopic redshift of $z = 0.062024$. We use pre-explosion template PS1/3$\pi$ images to determine the centroid of the host galaxy, which we find to be RA=15$^{h}$21$^{m}$20.087$^{s}$, Dec=+14$^{\circ}$04$^{'}$10.665$^{''}$ (with a centroid uncertainty of 0.09${''}$). To determine the location of AT\,2020neh in the pre-explosion reference images we performed astrometry using both YSE PS1 images of the transient close to maximum light in the $g$,$r$ and $i$ bands, and with acquisition imaging from Gemini from 11 Jul 2020. In all cases we use routine {\sc{iraf}} tasks to determine the coordinates of cross-matched point sources in both transient and reference images, then map and transform these coordinates to find the transient location in the reference image. We find a mean location of RA=15$^{h}$21$^{m}$20.094$^{h}$ Dec=+14$^{\circ}$04$^{'}$10.723$^{''}$ ($\pm$0.12\arcsec). We cross check this location using the late time ultraviolet imaging from HST. We find the HST location to be RA=15$^{h}$21$^{m}$20.089$^{h}$ Dec=+14$^{\circ}$04$^{'}$10.546$^{''}$ ($\pm$0.3\arcsec), consistent with our ground-based measurements. The larger uncertainty in the HST measurement is due to a lack of corresponding point sources in the ultraviolet image, so we use our ground based estimates as our final location. The location corresponds to a nuclear separation of 0.13\arcsec (or a physical separation of 0.16 $\pm$0.15 kpc), consistent with the transient being of nuclear origin (Figure 1). This separation is also consistent with the range of offsets for radio detected black holes in dwarf galaxies \citepmethods{Reines2020}. The nuclear location of the transient is inconsistent with the distribution of locations and offsets observed within the FBOT population from their apparent host galaxies \citepmethods[][]{Pursiainen2018,Wiseman2020}, which are not typically associated with the centres of their hosts. 

\section*{Host Galaxy Properties}

\subsection{\textit{Host-Galaxy Spectral Properties}}
We take optical spectroscopy of the host of AT\,2020neh from SDSS. We measure the emission lines present using standard routines within {\sc iraf} and plot the resulting emission line ratios on a BPT diagram \citepmethods[][]{Baldwin1981}, shown in Extended Data Figure 4. The BPT diagram uses emission line ratios to create theoretical regions used to identify the main excitation mechanisms which produce them \citepmethods{Kewley2006}; namely star formation, AGN shocks or a composite of the two. The host galaxy of AT\,2020neh lies within the star-forming region, close to the composite border. 

\subsection{\textit{Pre-explosion Variability}} 

Given the proximity of the host galaxy to the composite region of the BPT diagram, we consider the possibility that AT2020neh may have occurred in an environment with a pre-existing accretion disk (and therefore may be the result of AGN activity). From our early monitoring of the field with YSE (coverage from 24 Mar 2020 to 17 Jun 2020 in $g, r, i$ down to m$\sim$21.5), we find no significant variability outside of a 1$\sigma$ range from the mean of the forced photometry at the transient location over this 3 month period. We check for long-term variability using pre-explosion photometry available from ZTF (coverage from 27 Mar 2018 to 20 Jun 2020 in $g, r$) and from the Catalina Real Time Survey (CRTS) which monitored the host galaxy from 6 Apr 2005 to 16 Jun 2013 in a clear filter. We then use the fitting software {\sc{qso fit}} \citepmethods[][]{Butler2011} to fit the pre-explosion photometry data sets to determine whether the light curves exhibit significant variability characteristic of an AGN.  {\sc{qso fit}} generates a model light curve by modeling each point given the previous detections and a model co-variance matrix for AGN variability, then assesses how well the best-fit damped random walk model describes the data. Variable AGN typically have a significance of variability, $\sigma_{var}$, greater than 2 \citep{Baldassare2020}. We find the significance that the host galaxy of AT\,2020neh is variable to be low ($\sigma_{var}$=0.06 and 1.19 from ZTF and CRTS respectively) and the significance that the source is an AGN to also be low ($\sigma_{QSO}$=2.02 and 1.53 from ZTF and CRTS, where the typical significance for an AGN to be would is $\sigma_{QSO}>$2, \citep{Baldassare2020}). Therefore it is unlikely that AT\,2020neh harbours an active AGN or that the transient is due to AGN activity, in agreement with the host-galaxy location on the BPT diagram.

\subsection{\textit{Host Galaxy Spectral Energy Distribution}}

To perform Spectral Energy Distribution (SED) fitting of the host galaxy of AT\,2020neh, we use archival photometry from the PS1 3$\pi$ catalog in the $g,r,i,z$ and $y$ filters, alongside $J,H,K$ measurements from the 2 Micron All Sky Survey \citepmethods[2MASS, ][]{Skrutskie2006}, W1-3 bands from the Wide-field Infrared Survey Explorer \citepmethods[WISE; ][]{Wright2010} and NUV and FUV photometry from the Galaxy Evolution Explorer \citepmethods[GALEX; ][]{Martin2005}. To account for the systematic uncertainties in the photometry and in the physical models being fit to the emission, we apply a 10\% error floor to all photometry, with a 30\% error floor to the WISE W3 photometry, to account for variations in silicate absorption features around 10$\mu$m \citepmethods{Leja2017}. 

We then derive the physical properties of the host galaxy by fitting the collated photometry using the stellar population synthesis models in {\sc{Prospector}} \citepmethods{Leja2017}, considering wide priors on stellar mass, metallicity and star-formation history. We create a model which includes the effects of stellar and nebular emission, metallicity, dust reprocessing, and an exponentially declining star-formation history. We sample the posterior using the Bayesian nested sampling code {\sc{dynesty}} \citepmethods{Speagle2020}. The galaxy SED and prospector fitting results are shown in Extended Data Figure 4. The model encapsulates the shape of the host-galaxy SED well. We find a stellar mass $\log(M_{\star}/M_{\odot})) = 9.57^{+0.02}_{-0.01}$ and a low star-formation rate of $SFR = 0.33^{+0.06}_{-0.06} M_{\odot} yr^{-1}$ and a metallicity of $\log(Z/Z_{\odot}) = 0.13^{+0.19}_{-0.17}$. To account for the uncertainties introduced from assumptions made within the SED fitting on the galaxy star formation history and dust attenuation laws, we inflate our uncertainties in the stellar mass estimate to $\log(\Delta M_{\star}/M_{\odot})) = 0.19$, in accordance with the average uncertainty in stellar masses fitted with an exponentially declining star formation history in {\sc{Prospector}} \citepmethods{Lower2020}. 

Using the derived stellar mass, we can estimate the mass of the central black hole using a known scaling relation for low-mass galaxies \citep{ReinesVolonteri2015};

\begin{equation}
\log{\left( \frac{M_{BH}}{M_{\odot}}\right)} = \alpha + \beta \log{\left( \frac{M_{*}}{10^{11}M_{\odot}} \right)}
\end{equation}

where $\alpha = 7.45 \pm 0.08$ and $\beta = 1.05 \pm 0.11$. Using this scaling relation we estimate the mass of the black hole to be $\log{M_{BH}/M_{\odot}} = 5.95 ^{+0.20}_{-0.20}$

\subsection{\textit{Host Galaxy Spectral Modelling}}

We use the late time (+74d rest-frame) spectrum of AT\,2020neh taken with the DEIMOS spectrograph on Keck to measure the line-of-sight stellar velocity dispersion of the host galaxy. We use regions of the spectrum which are located away from the TDE line emission at this epoch to measure the stellar velocity dispersion to minimise contamination. The resolution of the DEIMOS spectrum (R$\approx$6000) is finer than that of the archival SDSS spectrum of the host galaxy, and given that the stellar velocity of dwarf galaxies is expected to be low \citep{Baldassare2020}, we chose to use this higher-resolution spectrum rather than the transient free SDSS spectrum to estimate the velocity dispersion. We use the Penalized Pixel Fitting software \citepmethods[pPXF; ][]{Cappellari2004,Cappellari2017} to determine the stellar kinematics of the galaxy. {\texttt{pPXF}} uses the spectral features in a galaxy spectrum alongside a set of stellar templates to determine the stellar line-of-sight velocity distribution.  We use the high-resolution spectral templates from the X-shooter Spectral Library \citepmethods{XSHlib2020}, which contains spectra of 628 stars in the UVB arm and 718 stars in the VIS arm at a spectral resolution of R$\approx 10,000$. We convolve these spectra with a Gaussian of $\sigma = 2.39 \AA$ to match them to the lower resolution of our DEIMOS spectrum (R$ \approx 6,000$). We focus on two wavelength regions covering the Mg Ib triplet at $5160 - 5190\AA$ and the Na-D doublet at $5890, 5895\AA$. We fit each of these regions separately, using the X-shooter UVB and VIS templates. We fit regions between $4700-5550\AA$ and $5400-6200\AA$ (both rest frame), masking the TDE spectrum outside of these windows, and fitting only for the first two moments of the line-of-sight velocity distribution (V and $\sigma$).

To estimate the uncertainty on the velocity dispersion fitted we use a Monte Carlo bootstrap method \citepmethods{Geha2009}, where we resample the 1-D spectrum based on the error spectrum, recalculate the velocity dispersion for 1000 noise realizations. From the resulting distribution of velocities we take the mean to be the final value, and the uncertainty to be the square root of the variance of the distribution. The best fitting stellar templates are presented in in Extended Data Figure 5. We find the independently measured velocity dispersions to be $\sigma=40\pm6$ km s$^{-1}$ and $\sigma=38\pm13$ km s$^{-1}$ fitting the UVB and VIS templates respectively. Given the low signal-to-noise of the host-galaxy stellar features in the VIS arm, we adopt our UVB measurement for our velocity dispersion estimate of the host galaxy.

From the velocity dispersion, we estimate the central black hole mass using using a $M_{BH}-\sigma$ relation \citepmethods{MerrittFerrarese2001};

\begin{equation} \label{eq1}
M_{BH} = 1.4 \times 10^{8}M_{\odot} \left( \frac{\sigma}{200 km s^{-1}}\right)^{4.72} .
\end{equation}

From our measured velocity dispersion, we estimate the mass of the black hole to be $\log{M_{BH}/M_{\odot}} = 4.8^{+0.4}_{-0.2}$.

\subsection{\textit{Stellar Surface Mass Density}}
We examine the stellar surface mass density, $\Sigma M_{*}$ of the host galaxy, which when considered alongside the velocity dispersion of the host galaxy, has been shown to be higher for TDE host galaxies compared to field galaxies in the local Universe \citep{Graur2018}, increasing the likelihood of a TDE occurring. The stellar mass surface density of the host galaxy is characterised as: 
\begin{equation}
\Sigma M_{*} = \frac{M_{*}/M_{\odot}}{r_{50}^{2}/kpc^{2}}
\end{equation}
where $M_{*}$ is our derived stellar mass and $r_{50}$ is the half-light radius enclosing 50\% of the host-galaxy flux. 
Using {\sc Source Extractor} we determine the 50\% flux radius using the pre-explosion template PS1/3$\pi$ image of the host galaxy in the SDSS r-band, setting a detection limit of 3$\sigma$ above the background sky. We measure an $r_{50}$ of 1.75 $\pm$ 0.07 kpc, which gives a surface mass density of $\log{\Sigma M_{*}}$=9.08.

\subsection{\textit{Comparison to TDE Host Galaxies}}

The host galaxy of AT2020neh has some similarities to the existing population of TDE hosts, but also has properties which make it unique.

The stellar mass of this galaxy places it towards the lower end of the TDE host-galaxy mass function, joining a small but growing number of optical TDEs in dwarf host galaxies \citepmethods[]{Hammerstein2022}. Though this and other measured properties of the host galaxy of AT\,2020neh (star formation rate and BPT location) are not significant outliers amongst other optical TDE host galaxies, \citep[][also see Extended Data Figure 4.]{French2020}, the color of the host galaxy is very blue ($u-r=1.7$), suggestive of on-going star formation. This is different to previous TDEs, whose higher mass host galaxies typically have low star formation rates or are post-starburst systems \citepmethods[][]{French2020,Hammerstein2021}, and does increase the possibility for contamination from SNe II. However, the host galaxy also possesses a relatively high S\'{e}rsic index ($n=3.2$) and stellar surface mass density compared to SDSS field galaxies. High S\'{e}rsic indices have been observed in other TDE hosts \citemethods{Graur2018,Hammerstein2021}. These over-densities of stars close to the galaxy nuclei are thought to enhance the TDE rate as they increasing the number of stars capable of being disrupted \citemethods[][]{French2020}. It may be that this is necessary for the production of a TDE within low mass starforming host galaxies. 

\section*{Light Curve Modelling}

The rate of fallback of the stellar debris from the disrupted star onto the central black hole and the peak timescale of the resultant TDE emission are sensitive to the mass of the disrupting black hole (\citemethods{Lodato2007,Guillochon2015,Mockler2019}). Assuming the emission from the TDE is \lq prompt\rq~ \citep[i.e. follows the fallback rate of material; ][]{Guillochon2015}, the light curve can be in principle used to measure the mass of the disrupting black hole. 

We fit the multi-band light curve of AT\,2020neh using the {\sc{MOSFiT}} light curve fitting code \citepmethods[Modular Open Source Fitter for Transients; ][]{Guillochon2018}. The {\sc{MOSFiT}} TDE model uses {\sc{FLASH}} hydrodynamic simulations of TDEs \citepmethods[][]{Guillochon2013} to calculate the fallback rate of debris for a given TDE set-up (i.e. black hole mass, stellar mass and impact parameter). Assuming a black body SED, which has shown to be an excellent approximation to TDE UV and optical emission \citep{vanVelzen2021}, this is then converted into a bolometric luminosity and parsed through reprocessing and viscosity transformation functions to generate multi-band light curves. Finally, {\sc{MOSFiT}} uses a Markov Chain Monte Carlo (MCMC) to fit the model light curves to the data. Full details of this model are presented within \citep{Mockler2019}. When fitting, we do not implement the default Eddington luminosity limit in the TDE model or limit the photospheric size of the transient. Most notably we are able to place firm constraints on the time of first fallback due to our early YSE photometric limits and detection. 

We present our light curve fits from all MCMC walkers in Extended Data Figure 6. We extract the posterior distributions for the key model parameters from {\sc{MOSFiT}}, and find AT\,2020neh is best described by models invoking a star of mass $M_{\star}=1.3^{+4.9}_{-1.0}M_{\odot}$, and impact parameter of $\beta=1.5^{+0.4}_{-0.9}$ (consistent with a near full disruption for the best-fit stellar mass) and a black hole with $\log{M_{BH}/M_{\odot}}=5.5^{+0.4}_{-0.3}$. The black hole mass is consistent with our previous estimates from the host-galaxy properties. We note that the mass of the star is somewhat degenerate with the efficiency parameter in this model \citep{Mockler2019}. To account for this, a systematic uncertainties of $\sigma_{\mathrm{M_{*}}}=0.66$ and $\sigma_{\mathrm{\epsilon}}=0.68$ are included in our error budget \citep{Mockler2019}. The strong nitrogen abundance observed in the early spectrum of AT\,2020neh without accompanying hydrogen could suggest a more evolved or stripped progenitor star, although high nitrogen abundances have been seen in other TDEs whose inferred stellar masses are consistent with the {\sc{MOSFiT}} stellar mass for 2020neh \citepmethods[][]{Mockler2021}.

The {\sc{MOSFiT}} BH mass is in good agreement with the BH masses inferred from two host galaxy measurements, increasing confidence that AT\,2020neh is produced by an IMBH candidate. However, there is still considerable debate as to the physical origin of the early optical and UV emission from TDEs, and it is uncertain how well they can be used as an independent way of measuring BH mass. This is needs to be carefully considered when using TDE inferred BH masses to constrain BH-galaxy scaling relationships. 

Though the BH masses inferred from light curve modelling have large inherent uncertainties due to the assumptions made within the model ({\sc{MOSFiT}} BH masses have a 0.2 dex systematic error), they are not always in agreement with masses derived via other methods \citep{Mockler2019}. For instance, some studies have found no correlation between TDE rise time and host-galaxy mass \citepmethods[][]{vanVelzen2021,Hammerstein2022}, which would be expected if the rise time follows the fallback timescale of material onto the BH. In particular, optical TDEs from other low mass host galaxies have been found to have generally higher {\sc{MOSFiT}} BH masses than those inferred from the host galaxy mass \citepmethods[]{Hammerstein2022}. However, alternative studies have found a moderate scaling between {\sc{MOSFiT}} BH masses and the mass of the host galaxy bulge \citemethods{Nicholl2022,Ramsden2022}. The discrepancy in BH mass estimates from TDE rise times may be due to the physical origin for the early optical/UV emission. If, rather than following the fallback of material, the rise time of a TDE were actually tracing the radiative diffusion timescale for photons to escape the shock-heated stellar debris, or if the photons were trapped by outflowing material, this would naturally lengthen the time taken to rise to peak \citemethods[]{vanVelzen2021, Gezari2021}. 

It is currently unclear which, if any, of the proposed physical mechanisms is prevalent in the production of early optical TDE emission. Though there is no firm consensus as to their utility for measuring BH masses, there is tentative evidence to suggest that they could be harnessed for this purpose, following a more sound understanding of their emission. Larger samples of well-observed TDEs across a range of galaxy masses are required to test this.


\section*{Rate estimations and predictions}

To estimate the local rate of TDEs with timescales and luminosities comparable to AT\,2020neh we assume the following; over 24 months of YSE operations we have observed only one AT\,2020neh-like event, monitoring fields for approximately 6 months each. This equates to one event per year within the YSE observational volume. At the time of writing, the observational footprint of YSE is 750 deg$^{2}$ of the northern hemisphere. We perform a volumetric correction to account for the maximum volume over which AT\,2020neh could have been detected within the YSE survey (V$_{\mathrm{max}}$). Under the current YSE observational strategy \citepmethods[][]{Jones2021}, we assume that a fast TDE may rise by 0.2 magnitudes before initial detection. This gives us a maximum redshift of $z=0.27$, which produces a volume of $\sim23\times 10^{6}$ Mpc$^{3}$. This provides an approximate observational rate of fast TDEs of $\lesssim 2 \times {\rm 10^{-8} Mpc^{-3} yr^{-1}}$ at $z=0.27$. 

For comparisons with the normal-TDE rate, we take the average per-galaxy TDE rate measurement of \citep{vanVelzen2020} for galaxies in the mass range $9.5 < \log{\frac{M_{*}}{M_{\odot}}}< 10.5$, and integrate this over the number of galaxies within our V$_{max}$ volume using the redshift dependant galaxy-mass functions of \citemethods{McLeod2021}, producing an average rate of $\sim {10^{-6.3} {\mathrm{Mpc}}^{-3} {\mathrm{yr}}^{-1}}$ at $z=0.27$. We emphasize that this is a conservative estimate as both stochastic effects from measuring rates based on a single object and small observing gaps that lower the YSE detection efficiency for fast transients could lower the estimated rate

The discovery and functionality of future fast TDE events will heavily depend upon the observing strategies of the surveys they are discovered in. Provided a survey has sufficient depth, future fast TDEs may be detected at any phase of their evolution. However, the utility of AT\,2020neh as a probe of the quiescent BH population lies in its early detection to constrain the rise time for BH mass estimates. Hence the cadence of the discovery survey will have the strongest impact on the usefulness of future fast TDEs. High-cadenced, deep surveys such as YSE will reliably identify fast TDEs in their infancy. With the planned doubling of the YSE footprint during 2022, based on our earlier rate estimate, we predict that YSE will identify an additional 5-6 events over the remaining 3 years of survey operations.

\section*{Data Availability Statement} All photometric data are available in Extended Data Table 3 and the spectra of AT\,2020neh will be made publically available to the community via WISeREP \citepmethods[][http://wiserep.org]{WISeREP2012}.

%
\bibliographystylemethods{naturemag}
\bibliographymethods{methods}

%
\end{methods}
%

%

\newpage
\begin{center}
    \leavevmode
    \includegraphics[width=1.0\hsize]{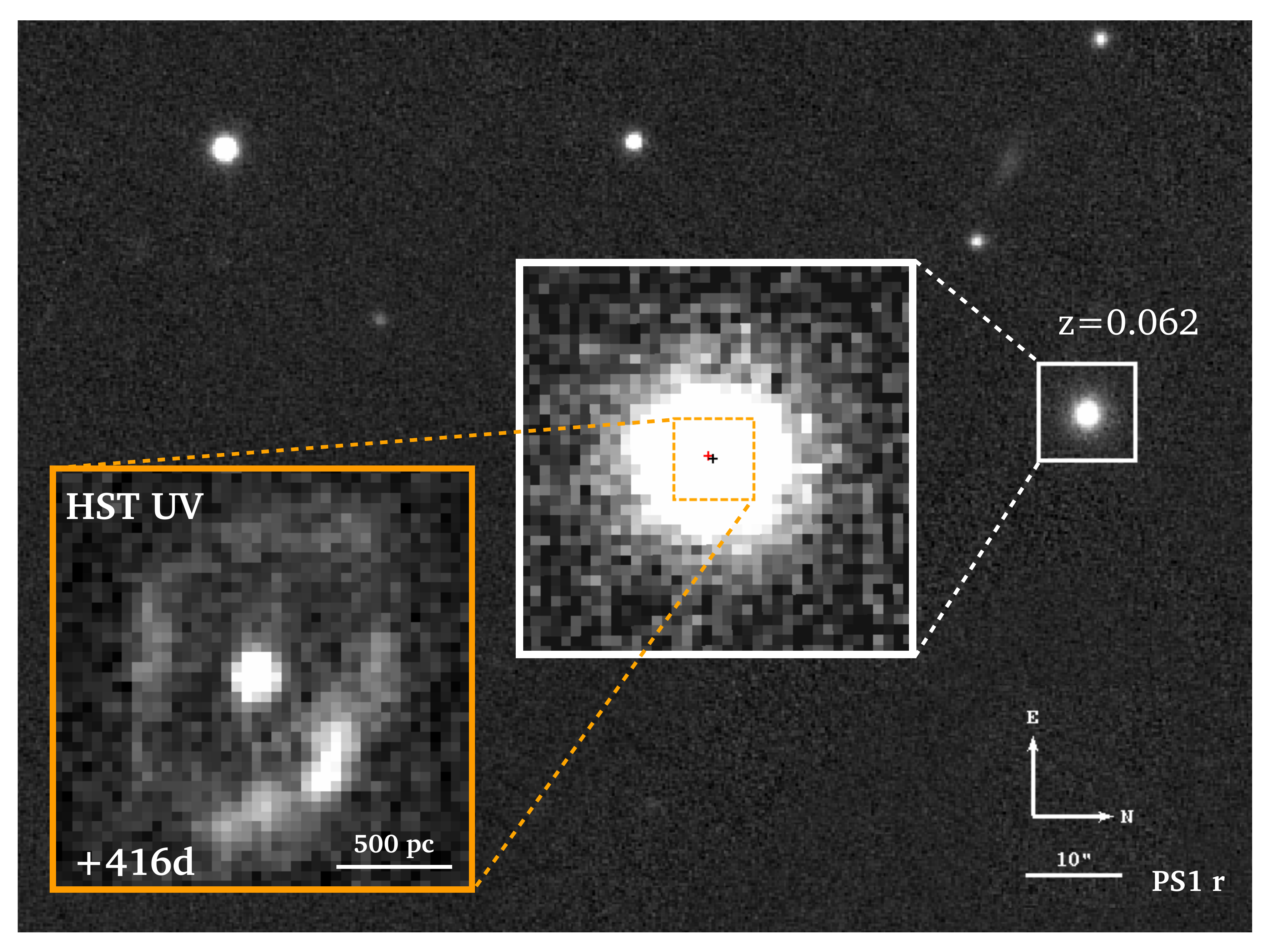}

\end{center}
{\noindent\bf Figure 1 $\mid$ The nuclear transient AT\,2020neh.} The wider image shows the environment of AT\,2020neh in optical PS1 $r$-band imaging. The dwarf host galaxy of the transient is highlighted. The first inset shows the apparent location of AT\,2020neh within its host galaxy in the optical. The host centroid is marked with a black cross whilst the location of AT\,2020neh is marked with a red cross (shown with 1$\sigma$ astrometric uncertainties). The location of the transient is coincident with the host nucleus. A second inset (orange boarder) shows deep ultraviolet imaging from the Hubble Space Telescope of AT\,2020neh at +416d. The transient is still clearly detected at the centre of the host, surrounded by a ring of star formation approximately 600pc from the nucleus. 

\newpage
%


%
\begin{center}
    \leavevmode
       \includegraphics[width=0.9\hsize]{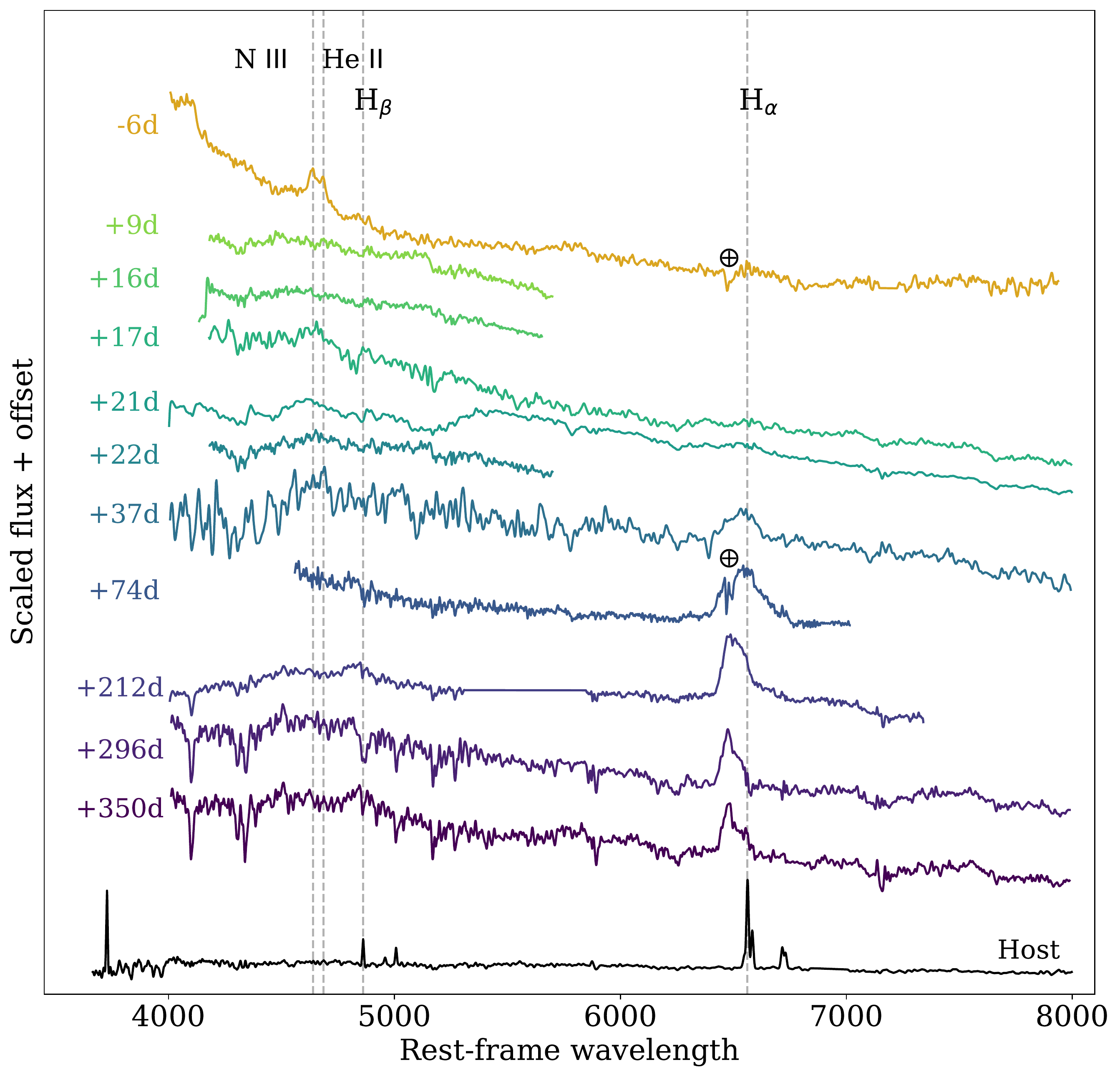}
\end{center}
{\noindent\bf Figure 2 $\mid$ 
The spectroscopic evolution of AT\,2020neh.} Common TDE emission features (H, He, N) are  marked. The strong He II and N III emission seen pre-maximum light disappears after the peak, with Balmer emission appearing at much later epochs, becoming increasingly asymmetric and blue-shifted as the TDE evolves. Spectra have been offset for clarity with rest-frame phases indicated. Crossed circles mark telluric features still present within the spectra. The instruments used to obtain this date are listed in Extended Data Table 1.

\begin{center}
    \leavevmode
    \epsfxsize=\textwidth
    \includegraphics[width=0.9\hsize]{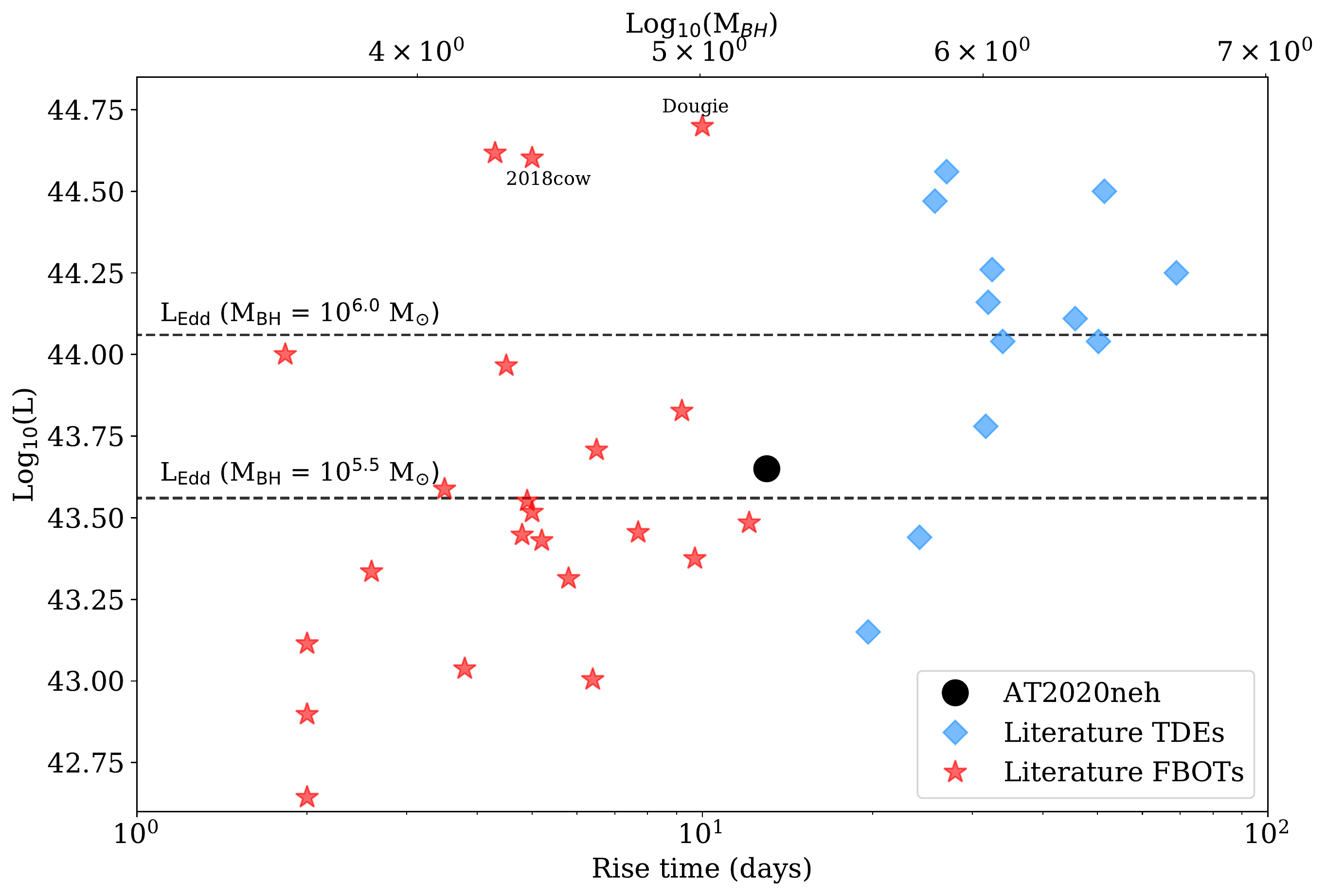}
\end{center}
 {\noindent\bf Figure 3 $\mid$ The rise-luminosity distribution of TDEs and FBOTs
}. The measured rise times of TDEs (blue circles) from the literature with pre-peak coverage as a function of their peak luminosities. AT\,2020neh occupies a unique position within this parameter space, being fast for a TDE and brighter than other short lived TDEs, lying between TDEs and the FBOT population, marked by red stars \citep{Pursiainen2018}. More energetic FBOT events such as \lq Dougie\rq~\citep{Vinko2015} and AT\,2018cow \citep{Prentice2018} are marked. Using the transient rise time to estimate the black hole mass for the TDE events (upper axis), we can see that \lq faster rising\rq~ AT\,2020neh like events probe the regions of super-Eddington accretion for lower (M$_{BH}<10^{5.5}$M$_{\odot}$) mass BHs (dashed lines). 
%
\newpage
\begin{center}
    \leavevmode
    \epsfxsize=\textwidth
    \includegraphics[width=1.0\hsize]{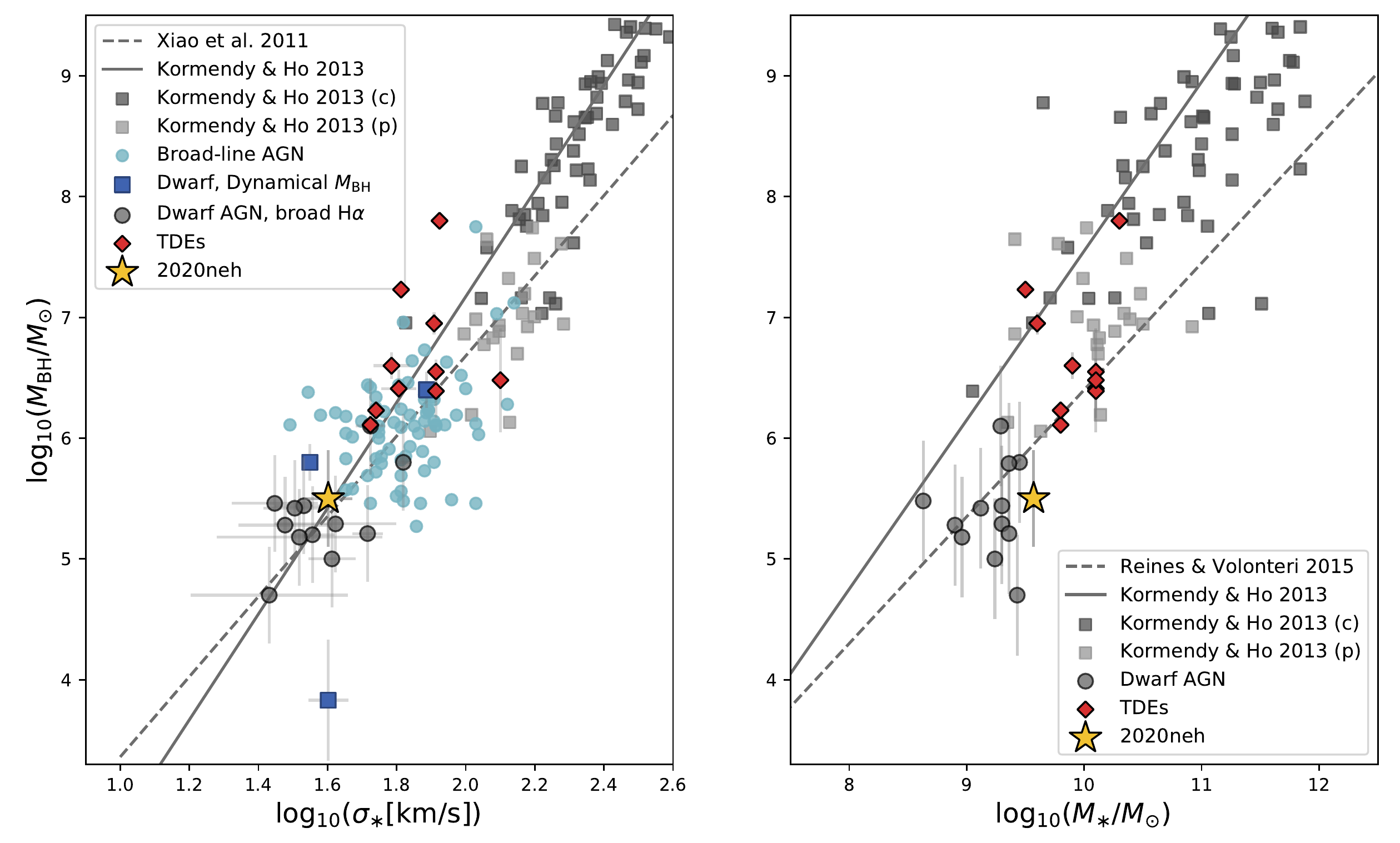}    
\end{center}
 {\noindent\bf Figure 4 $\mid$ Black hole scaling relations including dwarf galaxies
}. BH mass versus stellar velocity dispersion for samples of local galaxies ({\textit{left}}), alongside BH mass versus galaxy stellar mass ({\textit{right}}). Compilation samples of galaxies from \cite{Kormendy2013} are shown in grey squares, and broad line AGN from \cite{Xiao2011} are shown in light blue, alongside fits to the scaling relations. Dwarf AGN from \cite{Baldassare2017,Baldassare2020} are shown as dark grey circles. TDEs with BH mass estimates from \cite{Mockler2019} are shown in red, and AT\,2020neh is shown as a yellow star on both plots. The majority of galaxies with BH mass estimates in the in the low stellar mass/low velocity dispersion regime are those derived from the velocity of gas in the broad line region using the H-alpha line, which has large uncertainties. AT\,2020neh is one of a few BHs with an independent mass estimate in this region.

\newpage
\begin{center}
    \leavevmode
    \includegraphics[width=1.0\hsize]{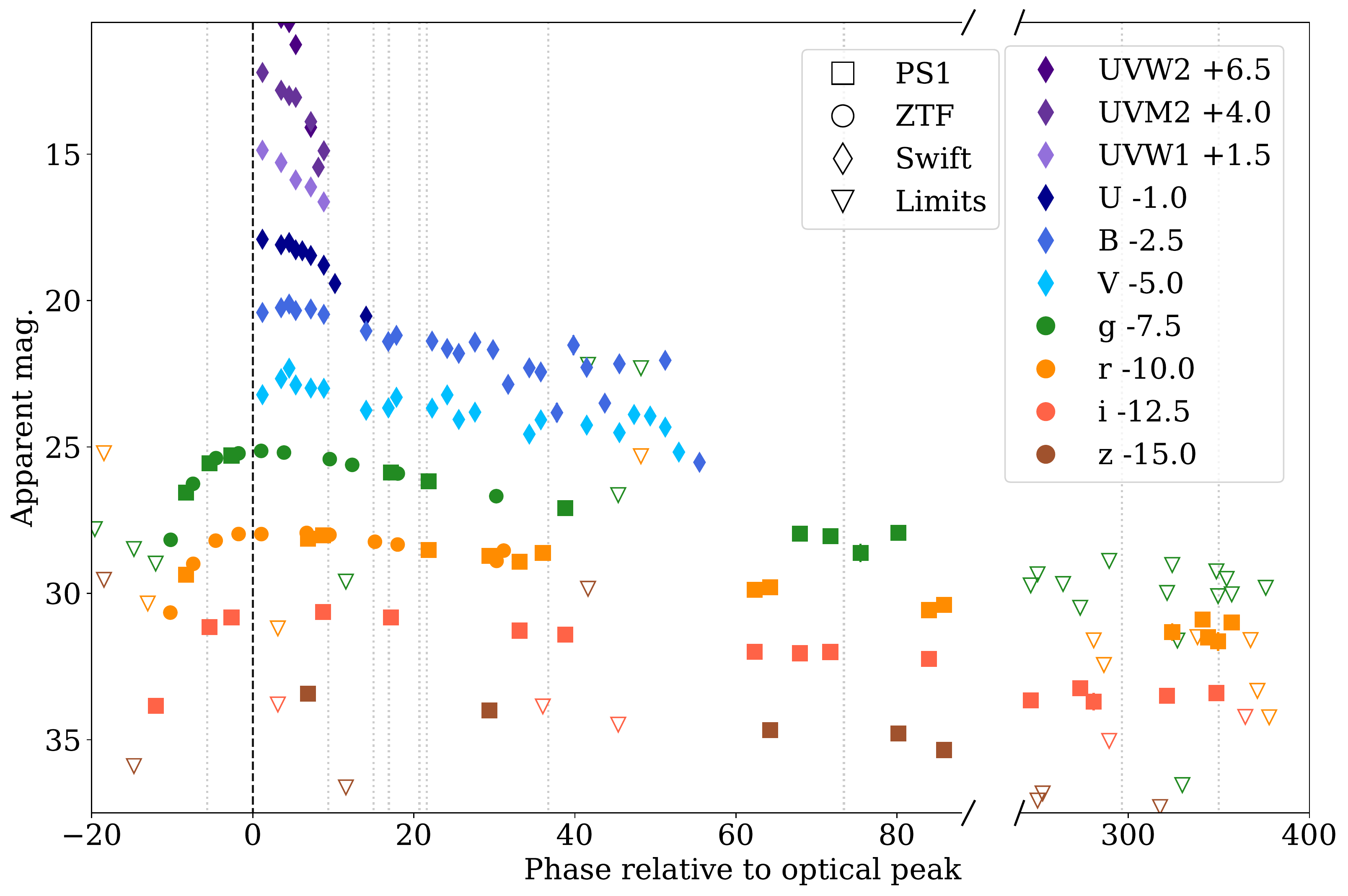}
\end{center}

{\noindent\bf Extended Data Figure 1 $\mid$ Light curve of AT\,2020neh.} The UV-optical light curve of AT\,2020neh throughout its observable lifetime. Grey dashed lines mark spectroscopic epochs, the dashed black line marks the peak of optical emission. Limits represent 3$\sigma$ upper limits to the flux for each telescope during periods of non-detection. There is an apparent gap in optical data between 40-60 days due to poor weather. AT\,2020neh is first detected within YSE (PS1) imaging in the $i$-band, confirmed by with a $g$-band detection from ZTF $<$24 hours later. Pre-explosion monitoring allows for strong constraints to be placed upon the explosion epoch of the transient.

%
\newpage
\begin{center}
    \leavevmode
    \includegraphics[width=0.9\hsize]{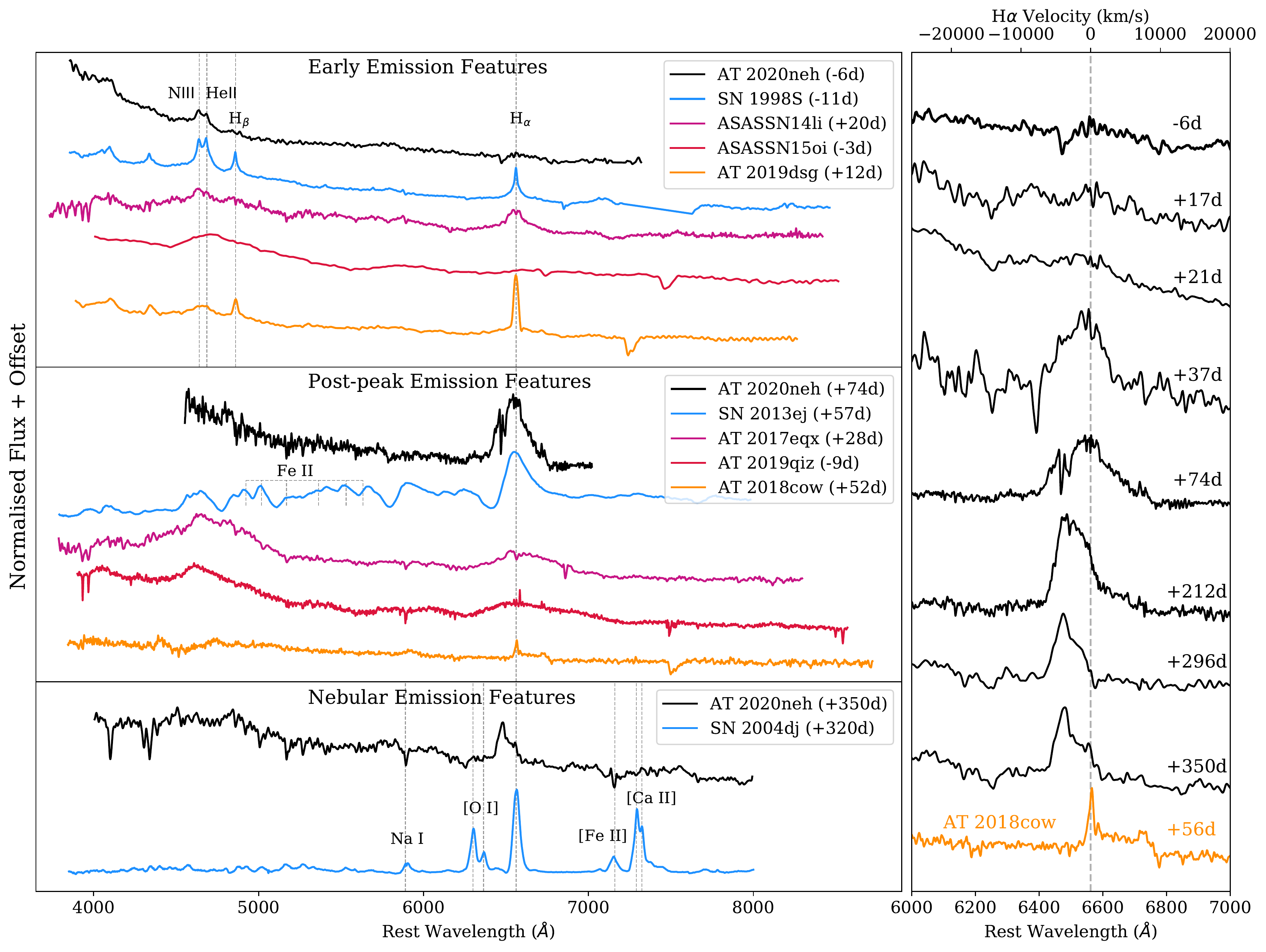}
\end{center}
{\noindent\bf Extended Data Figure 2 $\mid$ 
Spectroscopic comparison with other transients.} {\it Top:} The early time spectrum of AT\,2020neh alongside other TDEs which show Bowen fluorescence and an example of an alternative to this interpretation (CCSN flash-ionisation, as exemplified in SN\,1998S). {\it Middle:} Comparison of the post-maximum features in AT\,2020neh with other TDEs, a CCSN and the FBOT AT\,2018cow. AT\,2020neh presents a H$\alpha$ profile in emission, lacking the typical P-Cygni profile of a SN\,II but much broader than the FBOT. The rest of the spectrum is featureless, lacking absorption features from iron-group elements which would be expected of a CCSN during the photospheric phase. {\it Bottom:} The \lq nebular\rq\, phase spectrum of AT\,2020neh alongside a nebular phase CCSN. Expected emission features from SN during this phase are marked. {\it Right:} The evolution of the H$\alpha$ profile of AT\,2020neh. This feature appears late in the post-peak evolution, and retains a strongly blue-shifted profile, unseen in normal SNe. The narrow profile of AT\,2018cow is shown for comparison.

\newpage
\begin{center}
\leavevmode
\includegraphics[width=1.0\hsize]{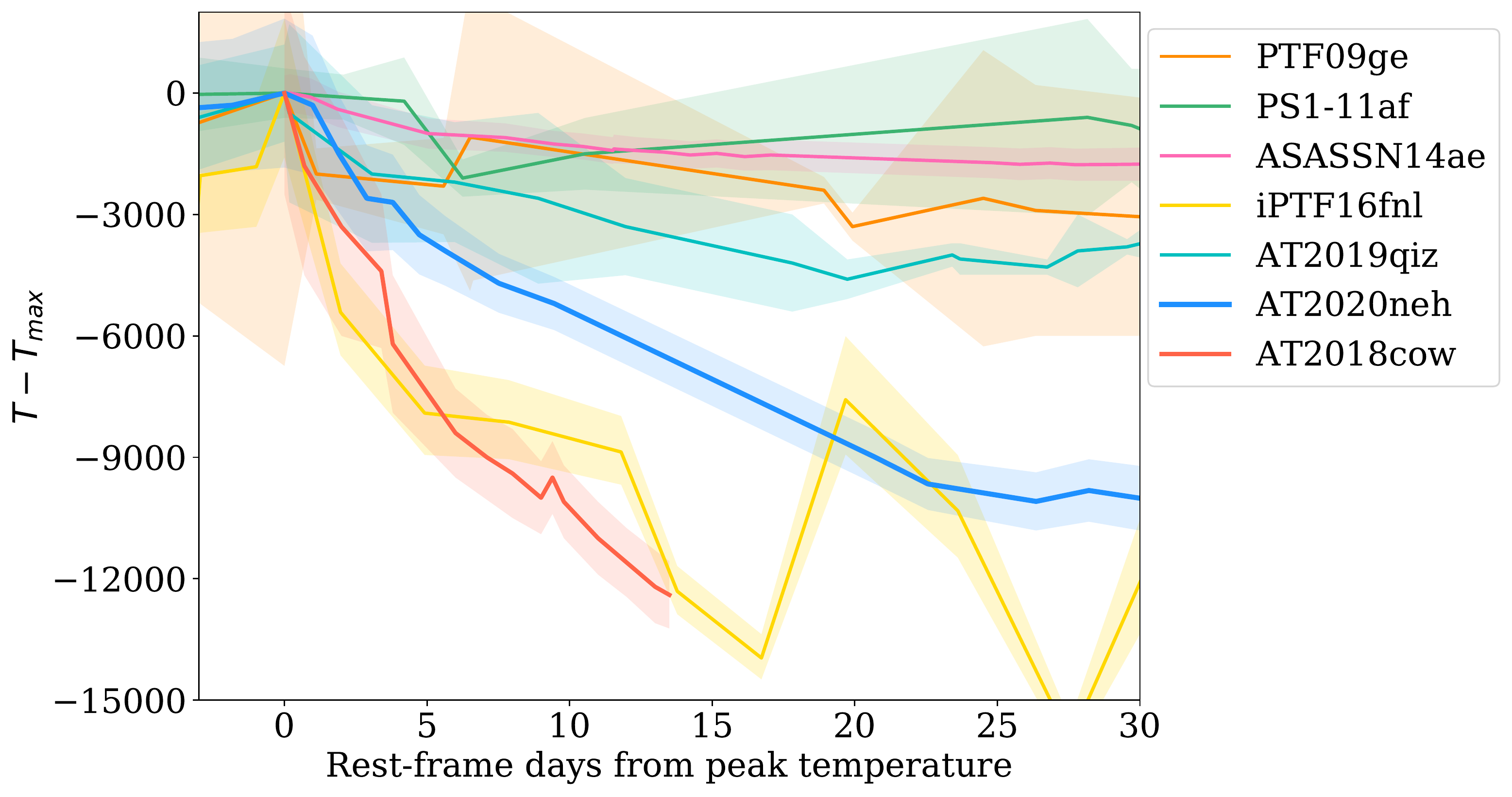}  
\end{center}
{{\noindent\bf Extended Data Figure 3 $\mid$ Cooling rates of transients} The early time cooling AT\,2020neh compared to that exhibited in the optically selected TDEs, including AT2019qiz and iPTF16fnl, alongside the FBOT AT\,2018cow. AT\,2020neh does exhibit significant cooling of approximately 9000k during the first 30 days from peak. This rate of cooling is comparable to that observed in iPTF16fnl and AT\,2019qiz, which are both believed to be produced by relatively low mass SMBHs (M$_{BH}\sim10^{6}  M_\odot$; \citemethods[][]{Blagorodnova2019,Nicholl2020}). This rate of cooling is much slower than that exhibited by AT\,2018cow, which cools by nearly 10,000K in the first 10 days from peak \citepmethods[]{Prentice2018}.}

\newpage
\begin{center}
\leavevmode
\includegraphics[width=1.0\hsize]{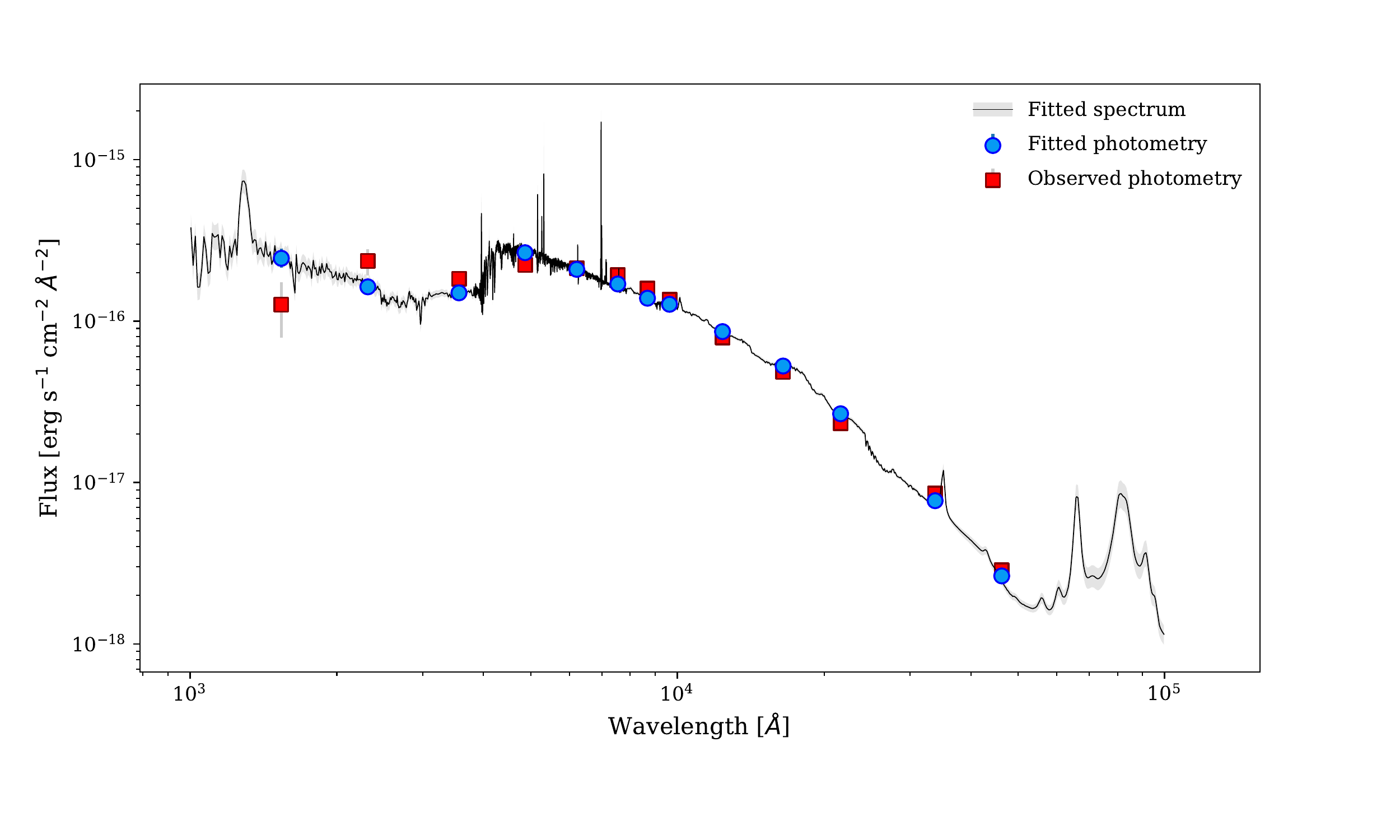} 
\includegraphics[width=1.0\hsize]{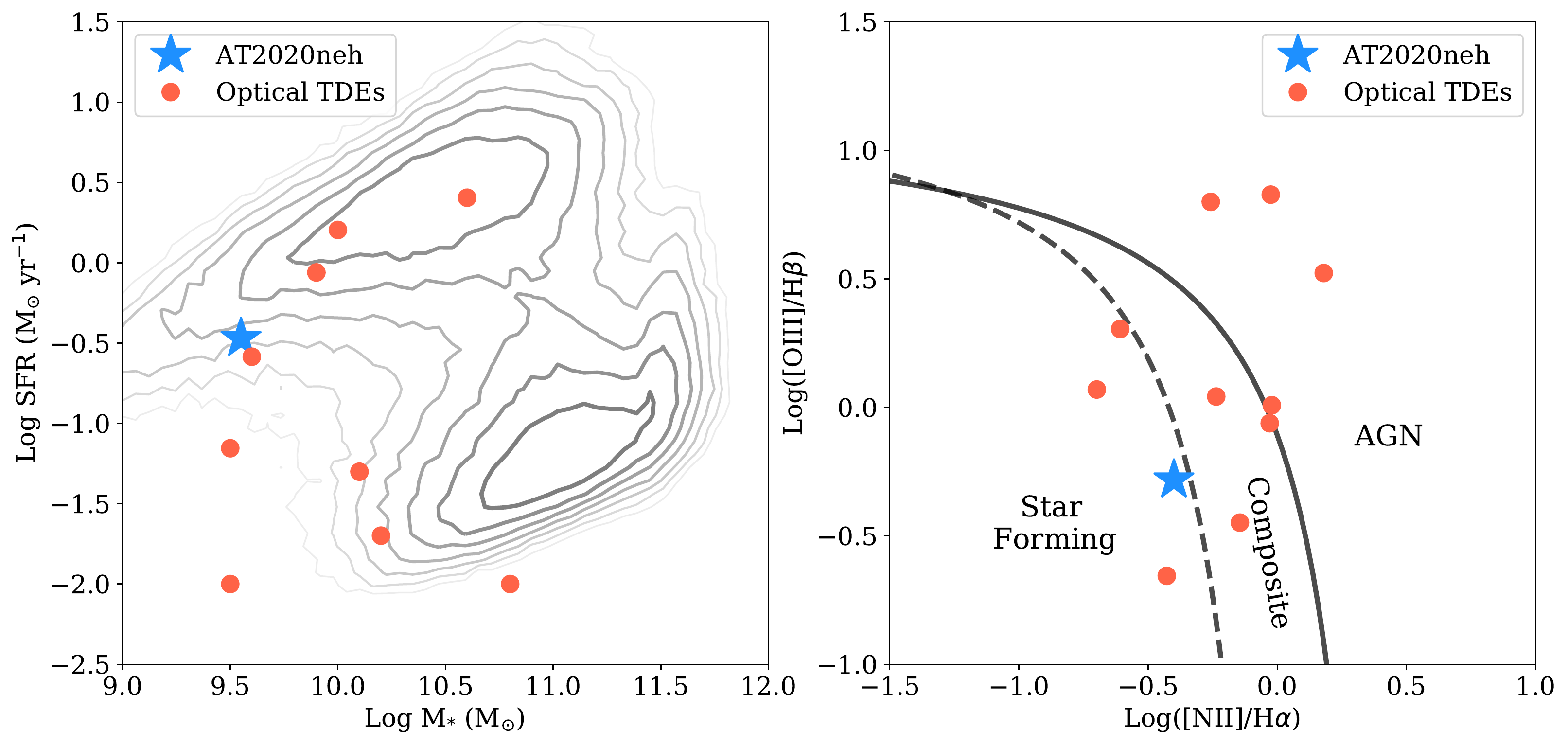}
\end{center}
{\noindent\bf Extended Data Figure 4 $\mid$ 
Properties of the host of AT\,2020neh} {\it Top:} The {\sc{Prospector}} SED fit to the host galaxy photometry {\it Bottom left:} Star formation rates vs. stellar masses for field galaxies (shown here is the SDSS spectroscopic sample as grey
contours). The hosts of optical TDEs are shown in red \citep[taken from][]{French2020}. AT\,2020neh sits comfortably within the parameter space occupied by other TDEs. {\it Bottom right:} A BPT diagram showing the regions where emission line ratios are indicative of ionisation
from either an AGN, star formation or a composite of the two \citep{Kewley2006}. The location of AT\,2020neh is shown, alongside other optical TDE hosts with host SDSS spectra \citep{French2020}. The position of AT\,2020neh inside of the star forming region of the diagram confirms the lack of an AGN component within the host.

\newpage
\begin{center}
    \leavevmode
    \includegraphics[width=1.\hsize]{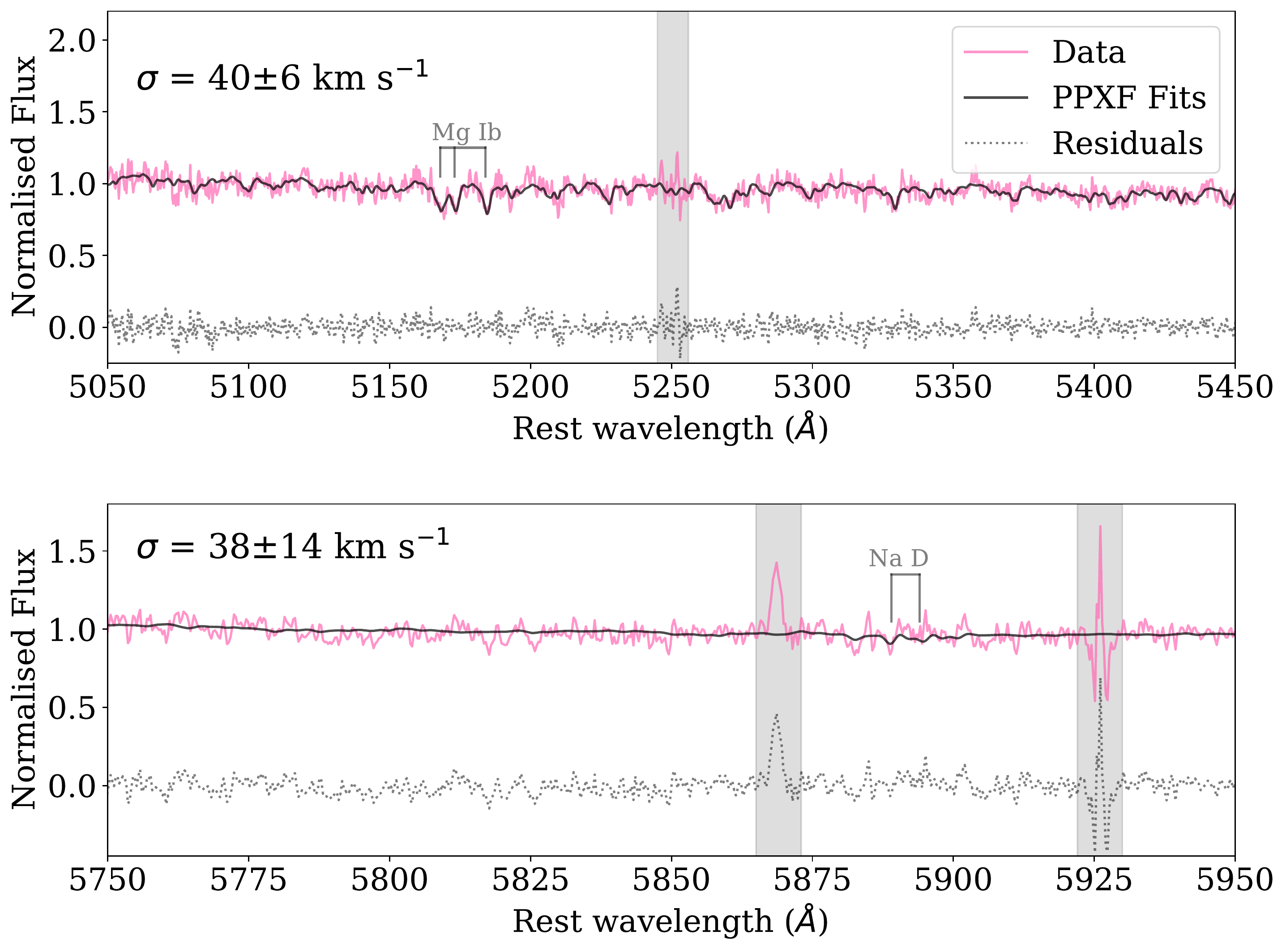}
\end{center}%
{\noindent\bf Extended Data Figure 5 $\mid$ Fitting the velocity dispersion of the host} The late-time DEIMOS spectrum of AT\,2020neh used to fit the stellar absorption features is shown, highlighting the regions around the Mg Ib triplet (top) and the Na D doublet (bottom). These two regions are fit independently. The best-fitting pPXF model is shown alongside each spectrum. We perform Monte-Carlo bootstrap fitting of the data to determine the uncertainties on the velocity dispersion. 
\newpage
\begin{center}
\leavevmode
\includegraphics[width=1.0\hsize]{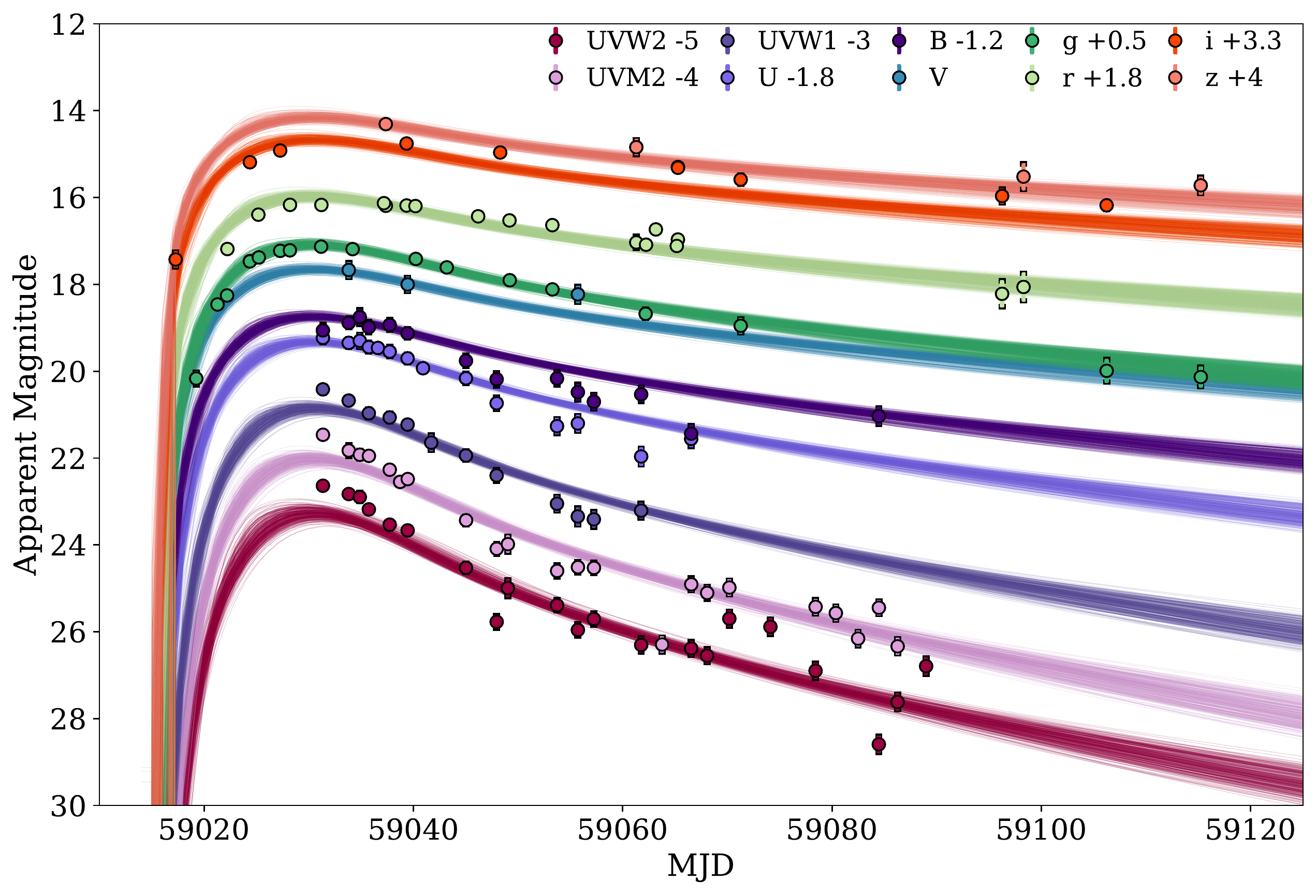}  
\end{center}
{\noindent\bf Extended Data Figure 6 $\mid$ 
{{\sc MOSFiT}} modelling of the light curve of AT\,2020neh} Photometry of AT\,2020neh in UV-optical bands (offset for clarity) alongside all possible TDE light curves each constructed from the posterior parameter distribution.

\newpage
\clearpage
\begin{table*}
\label{tab:LINLIST}
\captionsetup{labelsep=space}
\caption{$\vert$\ \ 
 {\bf H$\alpha$ Emission Properties} }
\centering
\footnotesize
\begin{tabular}{ccccc}
\hline
\noalign{\smallskip}
Phase        &
Flux  &
$\lambda_{observed}$  & 
Velocity offset &
EW \\
&($\times 10^{-15}$ erg s$^{-1}$)&($\AA$)&(km s$^{-1}$)&($\AA$)\\
\hline

\noalign{\smallskip}
+37d & 4.78 $\pm$ 0.04  &  6528 $\pm$ 3 &  -1526 $\pm$ 137 & 80 $\pm$ 9 \\
+74d & 16.10 $\pm$ 0.08  & 6525 $\pm$ 10 &  -1709 $\pm$ 457 &  105 $\pm$ 12 \\
+212d & 18.65 $\pm$ 0.02 &  6480 $\pm$ 1.0 & -3767 $\pm$ 46 & 129 $\pm$ 8\\
+296d & 3.02 $\pm$ 0.03  & 6475.1 $\pm$ 1.3 & -3990 $\pm$ 65 & 80 $\pm$ 3\\
+350d & 3.14 $\pm$ 0.04  & 6475.2 $\pm$ 1.7 & -3990 $\pm$ 79 & 82 $\pm$ 4\\
\hline
\end{tabular}\\
\end{table*}

\clearpage
\begin{table*}
\label{tab:LINLIST}
\captionsetup{labelsep=space}
\caption{$\vert$\ \ 
 {\bf Spectroscopic Followup Observations.} }
\centering
\footnotesize
\begin{tabular}{lcllll}
\hline
\noalign{\smallskip}
Date (UT) &
Phase        &
Telescope  &
Instrument   & 
Grating &
$\lambda$ Range ($\AA$) 
\\
\hline
\noalign{\smallskip}
25/06/2020  & -6 d & NOT & ALFOSC & Gr 4 &  4000 - 9000 \\ 
11/07/2020  & +9 d & Gemini & GMOS-N  & B1200 & 4400 - 6000 \\ 
17/07/2020  & + 15 d & Gemini & GMOS-N  & B1200 & 4400 - 6000 \\ 
19/07/2020  & +17 d & Shane-3m & KAST  & B452/3306 & 3300 - 6200 \\ 
& & & &R300/7500 & 3300 - 6200 \\ 
23/07/2020  & +21 d & Keck & LRIS & 400/3400 & 3200 - 5700   \\ 
 &  & &  & 400/8500 & 5500 - 10300   \\ 
24/07/2020  & + 22 d & Gemini & GMOS-N  & B1200 & 4400 - 6000 \\ 
09/08/2020  & +37 d & Shane-3m & KAST  & B452/3306 & 3300 - 6200 \\ 
& & & &R300/7500 & 3300 - 6200 \\ 
17/09/2020  & +74 d & Keck & DEIMOS & B1200 & 4800 - 7400  \\ 
12/02/2021  & +212 d & Keck & LRIS  & 400/3400 & 3200 - 5700   \\ 
 &  & &  & 400/8500 & 5500 - 10300   \\ 
12/05/2021  & +296 d & Keck & LRIS  & 400/3400 & 3200 - 5700   \\ 
 &  & &  & 400/8500 & 5500 - 10300   \\ 
08/07/2021  & +350 d & Keck & LRIS  & 400/3400 & 3200 - 5700   \\ 
 &  & &  & 400/8500 & 5500 - 10300   \\ 
\hline
\end{tabular}\\
\end{table*}

\clearpage
\begin{table*}
\label{tab:LINLIST}
\captionsetup{labelsep=space}
\caption{$\vert$\ \ 
 {\bf Photometry of AT~2020neh} }
\centering
\begin{tabular}{ c c }   
    \begin{tabular}{|cccc|}
    \hline
    \noalign{\smallskip}
    MJD       & Mag  & Mag Err  &  Filter \\
    \hline
    \noalign{\smallskip}
    59031.357 & 17.431 & 0.112 & U \\
    59033.821 & 17.546 & 0.096 & U \\
    59034.879 & 17.500 & 0.148 & U \\
    59035.744 & 17.644 & 0.112 & U \\
    59036.613 & 17.663 & 0.042 & U \\
    59037.736 & 17.748 & 0.116 & U \\
    59039.448 & 17.903 & 0.096 & U \\
    59040.923 & 18.131 & 0.042 & U \\
    59045.026 & 18.362 & 0.116 & U \\
    59047.956 & 18.937 & 0.145 & U \\
    59053.724 & 19.464 & 0.171 & U \\
    59055.713 & 19.401 & 0.181 & U \\
    59061.763 & 20.163 & 0.191 & U \\
    59066.541 & 19.758 & 0.177 & U \\
    59031.357 & 17.860 & 0.144 & B \\
    59033.822 & 17.686 & 0.111 & B \\
    59034.879 & 17.553 & 0.168 & B \\
    59035.745 & 17.787 & 0.129 & B \\
    59037.736 & 17.732 & 0.127 & B \\
    \hline
\end{tabular} &  
    \begin{tabular}{|cccc|}
    \hline
    \noalign{\smallskip}
    MJD       & Mag  & Mag Err  &  Filter \\
    \hline
    \noalign{\smallskip}
    59039.449 & 17.928 & 0.105 & B \\
    59045.027 & 18.558 & 0.131 & B \\
    59047.956 & 18.984 & 0.153 & B \\
    59053.724 & 18.966 & 0.154 & B \\
    59055.714 & 19.282 & 0.179 & B \\
    59057.244 & 19.501 & 0.169 & B \\
    59061.764 & 19.332 & 0.165 & B \\
    59066.542 & 20.239 & 0.189 & B \\
    59084.468 & 19.834 & 0.191 & B \\
    59033.826 & 17.668 & 0.171 & V \\
    59039.455 & 18.001 & 0.161 & V \\
    59055.720 & 18.229 & 0.183 & V \\
    59031.359 & 17.637 & 0.072 & UVW2 \\
    59033.824 & 17.828 & 0.065 & UVW2 \\
    59034.880 & 17.897 & 0.102 & UVW2 \\
    59035.746 & 18.181 & 0.081 & UVW2 \\
    59037.738 & 18.533 & 0.092 & UVW2 \\
    59039.452 & 18.665 & 0.076 & UVW2 \\
    59045.030 & 19.527 & 0.102 & UVW2 \\

\hline
\end{tabular} \\
\end{tabular}
\end{table*}

\begin{table*}
\centering
\begin{tabular}{ c c }   
    \begin{tabular}{|cccc|}
    \hline
    \noalign{\smallskip}
    MJD       & Mag  & Mag Err  &  Filter \\
    \hline
    \noalign{\smallskip}
    59047.959 & 20.771 & 0.144 & UVW2 \\
    59049.026 & 19.995 & 0.193 & UVW2 \\
    59053.727 & 20.389 & 0.131 & UVW2 \\
    59055.717 & 20.958 & 0.146 & UVW2 \\
    59057.247 & 20.706 & 0.143 & UVW2 \\
    59061.767 & 21.303 & 0.160 & UVW2 \\
    59066.544 & 21.384 & 0.158 & UVW2 \\
    59068.064 & 21.550 & 0.160 & UVW2 \\
    59070.188 & 20.697 & 0.171 & UVW2 \\
    59074.114 & 20.885 & 0.177 & UVW2 \\
    59078.425 & 21.898 & 0.174 & UVW2 \\
    59084.471 & 23.592 & 0.187 & UVW2 \\
    59086.263 & 22.616 & 0.178 & UVW2 \\
    59088.989 & 21.792 & 0.184 & UVW2 \\
    59031.363 & 17.463 & 0.066 & UVM2 \\
    59033.827 & 17.826 & 0.135 & UVM2 \\
    59034.882 & 17.922 & 0.101 & UVM2 \\
    59035.751 & 17.949 & 0.073 & UVM2 \\
    59037.743 & 18.268 & 0.083 & UVM2 \\
    59038.735 & 18.548 & 0.063 & UVM2 \\

    \hline
\end{tabular} 
\begin{tabular}{|cccc|}
    \hline
    \noalign{\smallskip}
    MJD       & Mag  & Mag Err  &  Filter \\
    \hline
    \noalign{\smallskip}
    59039.460 & 18.483 & 0.071 & UVM2 \\
    59045.037 & 19.433 & 0.100 & UVM2 \\
    59047.965 & 20.088 & 0.124 & UVM2 \\
    59049.029 & 19.982 & 0.180 & UVM2 \\
    59053.734 & 20.598 & 0.136 & UVM2 \\
    59055.724 & 20.512 & 0.128 & UVM2 \\
    59057.253 & 20.529 & 0.132 & UVM2 \\
    59063.765 & 22.293 & 0.170 & UVM2 \\
    59066.551 & 20.907 & 0.147 & UVM2 \\
    59068.071 & 21.105 & 0.146 & UVM2 \\
    59070.193 & 20.981 & 0.165 & UVM2 \\
    59078.431 & 21.426 & 0.164 & UVM2 \\
    59080.365 & 21.569 & 0.162 & UVM2 \\
    59082.490 & 22.157 & 0.164 & UVM2 \\
    59084.477 & 21.445 & 0.155 & UVM2 \\
    59086.269 & 22.335 & 0.174 & UVM2 \\
    59031.356 & 17.417 & 0.085 & UVW1 \\
    59033.819 & 17.675 & 0.078 & UVW1 \\
    59035.743 & 17.967 & 0.097 & UVW1 \\
    59037.735 & 18.064 & 0.100 & UVW1 \\

    \hline
\end{tabular} 
\end{tabular} 
\end{table*}

\begin{table*}
\centering
\begin{tabular}{ c c }   
    \begin{tabular}{|cccc|}
    \hline
    \noalign{\smallskip}
    MJD       & Mag  & Mag Err  &  Filter \\
    \hline
    \noalign{\smallskip}
    59039.446 & 18.228 & 0.084 & UVW1 \\
    59041.708 & 18.644 & 0.177 & UVW1 \\
    59045.024 & 18.941 & 0.110 & UVW1 \\
    59047.954 & 19.400 & 0.139 & UVW1 \\
    59053.722 & 20.051 & 0.166 & UVW1 \\
    59055.711 & 20.343 & 0.193 & UVW1 \\
    59057.241 & 20.411 & 0.179 & UVW1 \\
    59061.762 & 20.210 & 0.166 & UVW1 \\
    58958.530 & 21.095 & 0.306 & PS1 g \\
    58968.426 & 21.241 & 0.254 & PS1 g \\
    58990.375 & 21.282 & 0.287 & PS1 g \\
    59021.287 & 18.962 & 0.051 & PS1 g \\
    59024.381 & 17.964 & 0.020 & PS1 g \\
    59027.278 & 17.727 & 0.026 & PS1 g \\
    59053.284 & 18.615 & 0.041 & PS1 g \\
    59071.274 & 19.453 & 0.163 & PS1 g \\
    59106.237 & 20.490 & 0.259 & PS1 g \\
    59115.231 & 20.633 & 0.225 & PS1 g \\
    59024.381 & 17.970 & 0.020 & PS1 g \\
    58958.533 & 20.786 & 0.201 & PS1 r \\
\hline
\end{tabular} 
\begin{tabular}{|cccc|}
    \hline
    \noalign{\smallskip}
    MJD       & Mag  & Mag Err  &  Filter \\
    \hline
    \noalign{\smallskip}
    58968.429 & 21.027 & 0.202 & PS1 r \\
    58997.342 & 21.215 & 0.251 & PS1 r \\
    59037.367 & 17.989 & 0.046 & PS1 r \\
    59039.352 & 17.983 & 0.030 & PS1 r \\
    59053.281 & 18.435 & 0.038 & PS1 r \\
    59061.299 & 18.833 & 0.140 & PS1 r \\
    59065.275 & 18.767 & 0.042 & PS1 r \\
    59096.249 & 20.017 & 0.298 & PS1 r \\
    59098.298 & 19.859 & 0.317 & PS1 r \\
    58942.491 & 20.369 & 0.358 & PS1 i \\
    58960.455 & 20.562 & 0.186 & PS1 i \\
    58966.516 & 20.771 & 0.327 & PS1 i \\
    58972.442 & 20.602 & 0.196 & PS1 i \\
    58990.372 & 20.522 & 0.151 & PS1 i \\
    59017.287 & 20.726 & 0.171 & PS1 i \\
    59024.383 & 18.487 & 0.034 & PS1 i \\
    59027.276 & 18.217 & 0.030 & PS1 i \\
    59039.349 & 18.055 & 0.036 & PS1 i \\
    59048.298 & 18.263 & 0.031 & PS1 i \\
    59065.278 & 18.614 & 0.046 & PS1 i \\
\hline
\end{tabular} 
\end{tabular} 
\end{table*}

\begin{table*}
\centering
\begin{tabular}{ c c }   
    \begin{tabular}{|cccc|}
    \hline
    \noalign{\smallskip}
    MJD       & Mag  & Mag Err  &  Filter \\
    \hline
    \noalign{\smallskip}
    59065.278 & 18.598 & 0.045 & PS1 i \\
    59071.271 & 18.886 & 0.103 & PS1 i \\
    59096.252 & 19.265 & 0.156 & PS1 i \\
    59106.234 & 19.479 & 0.092 & PS1 i \\
    59037.364 & 18.308 & 0.079 & PS1 z \\
    59061.302 & 18.840 & 0.170 & PS1 z \\
    59098.295 & 19.519 & 0.294 & PS1 z \\
    59115.228 & 19.719 & 0.187 & PS1 z \\
    59019.255 & 20.667 & 0.145 & ZTF g \\
    59022.199 & 18.755 & 0.032 & ZTF g \\
    59025.231 & 17.879 & 0.021 & ZTF g \\
    59028.201 & 17.716 & 0.020 & ZTF g \\
    59031.189 & 17.631 & 0.029 & ZTF g \\
    59034.195 & 17.688 & 0.030 & ZTF g \\
    59040.232 & 17.913 & 0.017 & ZTF g \\

\hline
\end{tabular} 
\begin{tabular}{|cccc|}
    \hline
    \noalign{\smallskip}
    MJD       & Mag  & Mag Err  &  Filter \\
    \hline
    \noalign{\smallskip}
    59043.191 & 18.108 & 0.019 & ZTF g \\
    59049.207 & 18.404 & 0.026 & ZTF g \\
    59062.191 & 19.177 & 0.103 & ZTF g \\
    59022.232 & 18.985 & 0.034 & ZTF r \\
    59025.190 & 18.194 & 0.026 & ZTF r \\
    59028.213 & 17.968 & 0.024 & ZTF r \\
    59031.211 & 17.970 & 0.043 & ZTF r \\
    59037.189 & 17.930 & 0.023 & ZTF r \\
    59040.197 & 17.996 & 0.019 & ZTF r \\
    59046.190 & 18.232 & 0.023 & ZTF r \\
    59049.185 & 18.326 & 0.026 & ZTF r \\
    59062.229 & 18.886 & 0.074 & ZTF r \\
    59063.168 & 18.534 & 0.044 & ZTF r \\
    59065.168 & 18.914 & 0.070 & ZTF r \\
& & & \\
\hline
\end{tabular} 
\end{tabular} 
\end{table*}

\clearpage
%
%
\end{document}